%% file: JADES_quiescent.tex
\documentclass[longauth]{aa}  
\usepackage{etoolbox} 
\usepackage{subcaption}
\usepackage{graphicx}
\usepackage{textgreek}
\usepackage{ulem}
\usepackage[svgnames]{xcolor}
\usepackage{booktabs}

\usepackage{txfonts}
\usepackage{hyperref}
\usepackage{xspace}
\hypersetup{
    colorlinks=true,
    linkcolor=blue,
    citecolor=blue,
    filecolor=magenta,      
    urlcolor=blue,
    pdftitle={JADES: environment-quenched galaxy at z=2.3},
    pdfpagemode=FullScreen,
    }
\makeatletter
\renewcommand*\aa@pageof{, page \thepage{} of \pageref*{LastPage}}
\makeatother

\newcommand{\kms}{km\,s$^{-1}$\xspace}

\usepackage{adjustbox}

\newcommand{\orcid}[2]{\href{http://orcid.org/#2}{#1{\includegraphics[height=10pt]{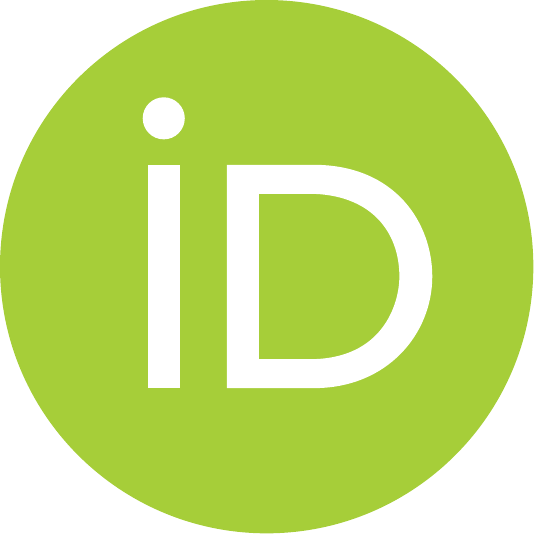}}}}

\newcommand{\Mstar}{\ensuremath{M_\star}\xspace}
\newcommand{\re}{\ensuremath{R_\mathrm{e}}\xspace}

\newcommand{\peryr}{\ensuremath{\mathrm{yr}^{-1}}\xspace}

\newcommand{\MSun}{\ensuremath{{\rm M}_\odot}\xspace}
\newcommand{\ZSun}{\ensuremath{{\rm Z}_\odot}\xspace}

\newcommand{\fluxcgs}{\ensuremath{\mathrm{erg\,s^{-1}\,cm^{-2}}}\xspace}
\newcommand{\mum}{\text{\textmu m}\xspace}
\let\oldAA\AA
\renewcommand{\AA}{\oldAA\xspace}
\let\oldtextsigma\textsigma
\renewcommand{\textsigma}{\oldtextsigma\xspace}

\newcommand{\targetidfull}{JADES-GS+53.12365-27.80454\xspace} 
\newcommand{\targetidcentral}{JADES-GS+53.12314-27.80346\xspace} 

\newcommand{\lya}{\text{Ly\textalpha}\xspace}
\newcommand{\oiii}{[O\,\textsc{iii}]\xspace}
\newcommand{\Hbeta}{\text{H\textbeta}\xspace}
\newcommand{\Halpha}{\text{H\textalpha}\xspace}
\newcommand{\Pabeta}{\text{Pa\textbeta}\xspace}

\newcommand{\tvdiff}{\ensuremath{\tau_{V,\,\mathrm{diff}}}\xspace}
\newcommand{\tvbc}{\ensuremath{\tau_{V,\,\mathrm{BC}}}\xspace}

\newcommand{\jwst}{\textit{JWST}\xspace}
\newcommand{\hst}{\textit{HST}\xspace}

\newcommand{\beagle}{{\sc beagle}\xspace}
\newcommand{\prospector}{{\sc prospector}\xspace}
\newcommand{\eazy}{{\sc eazy}\xspace}
\newcommand{\forcepho}{{\sc forcepho}\xspace}

\newcommand{\tauV}{\hbox{$\hat{\tau}_\textsc{v}$}}
\newcommand{\Us}{\hbox{$U_\textsc{s}$}}
\newcommand{\logUs}{\hbox{$\log\Us$}}
\newcommand{\xid}{\hbox{$\xi_\mathrm{d}$}}
\newcommand{\mud}{\hbox{$\mu_\mathrm{d}$}}

\newcommand{\zfinal}{2.34\xspace} 
\newcommand{\mfinal}{\ensuremath{9.5^{+1.8}_{-1.2}\times10^8}\xspace}
\newcommand{\logmfinal}{8.98\xspace} 
\newcommand{\agefinal}{0.8\xspace} 
\newcommand{\zquench}{2.9\xspace}
\newcommand{\quenchduration}{0.6\xspace} 
\newcommand{\tquench}{2.2\xspace} 

\begin{document} 

   \title{JADES: deep spectroscopy of a low-mass galaxy at redshift 2.3 quenched by environment}

    \titlerunning{JADES: environment-quenched galaxy at z=2.3}

    \input{authors}

   \authorrunning{L. Sandles et al.}
   \date{}

  \abstract
{
  We report the discovery of a quiescent galaxy at $z=2.34$ with a stellar mass of only $M_\star = \mfinal$~\MSun, based on deep \jwst/NIRSpec spectroscopy. This is the least massive quiescent galaxy found so far at high redshift.
  We use a Bayesian approach to model the
  spectrum and photometry, and find the target to have been quiescent for \quenchduration~Gyr with a mass-weighted average stellar age of \agefinal--1.7~Gyr (dominated by systematics). The galaxy displays an inverse colour
  gradient with radius, consistent with environment-driven
  quenching.
  Based on a combination of spectroscopic and
  robust (medium- and broad-band) photometric redshifts, we identify a galaxy
  overdensity near the location of the target (5-\textsigma above the background level
  at this redshift). We stress that had we been specifically targetting galaxies within
  overdensities, the main target would not have been selected on photometry alone; therefore,
  environment studies based on photometric redshifts are biased against low-mass
  quiescent galaxies.
  The overdensity contains three
  spectroscopically confirmed, massive, old galaxies ($M_\star =
  8\text{--}17\times10^{10}$~\MSun). The presence of these evolved systems points to
  accelerated galaxy evolution in overdensities at redshifts $z>2$, in agreement with previous
  works. In projection, our target lies only 35~pkpc away from the most 
  massive galaxy in this overdensity (spectroscopic redshift $z=2.349$) which is located close to overdensity's centre.
  This suggests the low-mass galaxy was quenched by environment, making it
  possibly the earliest evidence for environment-driven quenching to date.
}

   \keywords{Galaxies: high-redshift, formation, evolution, stellar content, star formation, statistics}

   \maketitle
%

\section{Introduction} \label{s.intro}

Recent observations have shown that the young, distant Universe was an eventful place,
with many -- or even most -- galaxies undergoing starbursts
\citep{endsley_jwst_nircam_2022, dressler_building_the_first_2023, looser_jades_differing_2023}. Most of these early
galaxies
must have moved on with their star-formation histories (SFH) to
more stable star-formation rates (SFR); but perhaps some were left
behind, permanently quenched by the excesses of feedback, or by their more
massive peers.

At any epoch we have been able to study, star-forming galaxies follow a relation between 
their SFR and their stellar mass -- the so-called star-forming main 
sequence \citep[SFMS; e.g.,][]{brinchmann+2004, noeske+2007, renzini+peng2015}.
While star formation in local galaxies appears regulated by local processes
\citep{wang+2019, Bluck2020a, baker+2022}, it is equally clear that global conditions also 
play a critical role \citep{tacchella_confinement_2016}. For example, the SFMS evolves with redshift \citep[e.g.,][]{Speagle_MS_14, Sandles_MS_22, Popesso_MS_23}, likely a response to the
increased availability and density of cold gas at earlier epochs \citep[e.g.,][]{tacconi_evolution_2020, baker+2023}.
At some point, however, star formation decreases and eventually stops, giving rise to
an inexorably increasing population of defunct galaxies -- without appreciable star 
formation in the last hundreds of Myr and longer (quiescent galaxies).
There are many mechanisms which can stop (quench) star formation, which we can classify
as `internal' or `external'. Internal mechanisms include ejective or preventative feedback from star formation \citep{white_rees_core_1978,dekel_silk_origin_1986, cole_hierachical_2000}
and supermassive black holes \citep{silk+rees1998,binney2004,Croton2006MNRAS.365...11C},
virial shocks \citep{dekel_cold_flows_shock_2006}, and gas stabilisation against fragmentation \citep{martig_morphological_2009, gensior_heart_2020}.
External mechanisms are gas ionisation due to background radiation \citep{efstathiou_suppressing_1992}, and gas removal, heating, non-accretion and overconsumption due to environmental effects,
\citetext{e.g., \citealp{gunn_gott_infall_1972, 
vogelsberger_illustris_2014, herniques_formation_2015, 
cortese_dawes_2021, wright_orbital_2022}; see 
\citealp{man_belli_quenching_2018} for a review of quenching mechanisms}.

Even though there is ample evidence for overdensities at high redshift \citetext{e.g., at $z=5\text{--}7$, \citealp{Lemaux_VIMOS_18, laporte_a_dense_2022,
li_groups_2022, brinch_cosmos2020_2023, helton_jades_2023}},
the role of environment in these early overdensities remains unclear, and may range
from enhanced star formation fed by cold-gas
streams \citep[e.g.,][]{Narayanan_Formation_2015, Umehata_Gas_2019} to enhanced quenched fractions of
ultra-massive galaxies \citetext{e.g., \citealp{chartab_largescale_2020, Kubo_Massive_2021, Shi_accelerated_21, McConachie_Spectroscopic_2022}; see \citealp{alberts_noble_2022} for a review}.

At the very massive end, some studies combining ground-based 
optical spectroscopy with photometry have found galaxies in 
overdensities at $z>2$ to show evidence for 
accelerated evolution \citep[e.g.,][]{Lemaux_VIMOS_14, 
Lemaux_VIMOS_18}, potentially low gas fractions 
\citep{zavala_gas_content_2019} and, above $\Mstar = 
3\times10^{10}$~\MSun, higher quiescent fractions 
\citep{shimakawa_mahalo_2018, Shi_accelerated_21}.
Drawing a line to environmental quenching mechanisms is 
challenging however: for one, massive, virialized clusters at 
$z\approx1.5$ -- expected to have a strong potential for 
environmental quenching given the galaxy density and 
established intra-cluster medium - have been shown to host 
vigorous field-like star formation, including in very massive 
galaxies \citep[i.e.,][]{Brodwin_Era_2013, Ma_Dusty_2015, Santos_Reversal_2015, Alberts_Star_2016, lemaux_vimos_2022}, and have mixed evidence for enhanced or depleted 
gas fractions \citetext{cf.~\citealp{zavala_gas_content_2019}
and \citealp{noble_alma_2017, alberts_significant_2022, 
williams_alma_2022}}.
Furthermore, at high mass, quenching is also driven by internal mechanisms \citep[as evidenced 
by the high fraction of massive, quiescent centrals, e.g.,][]{peng+2010, donnari_quenched_2021}.
Therefore, if we are to unambiguously study environment-driven quenching, we must probe
the low-mass end of the galaxy distribution, in the mass regime $\Mstar = 
10^8\text{--}10^{10}$~\MSun where galaxies are too massive for re-ionisation quenching
\citep[e.g.,][]{Ma2018MNRAS.478.1694M},
but not massive enough for internal quenching \citep{bluck+2020b,donnari_quenched_2021}.
\citet{ji_enviromental_2018} have used a sample of colour-selected galaxies to show that 
quiescent galaxies are more clustered than coeval star-forming galaxies at redshifts $z=1.5\text{--}2.5$. In particular, they find some evidence for low-mass quiescent galaxies 
being more clustered than high-mass quiescent galaxies, which is expected from
environment-driven quenching. However, until now, spectroscopic evidence for 
quiescent, low-mass satellites is still missing.

\jwst is revolutionising our understanding of galaxy quenching.
Besides discovering an abundance of starburst galaxies at early cosmic epochs, \jwst also
betrayed an overabundance of high-mass quiescent galaxies \citep{carnall+2023b},
out to $z>4.5$ \citep{carnall+2023c}. These galaxies are defined by having low specific SFR (sSFR) in the last 100~Myr prior to observation \citep[e.g.,][]{pacifici+2016}. These findings challenge current models
of quenching, which predict smaller fractions -- particularly for high-mass galaxies
at $z>3$ \citep[e.g.,][]{valentino_atlas_2023}.
This discrepancy suggests that feedback in the young Universe was more powerful and/or
feedback/gas consumption were more efficient than previously thought, a picture that is also supported empirically \citep{Whitaker2021Natur.597..485W, williams_alma_2021, Suzuki2022}. At the same time, 
lower-mass quiescent galaxies ($\Mstar = 10^{10}$~\MSun) have
been spectroscopically confirmed at redshifts as high as $z=2\text{--}2.5$ 
\citep{Marchesini2023ApJ...942L..25M},
and even lower-mass, rapidly quenched systems up to $z=5$ \citep{Strait2023} and
$z=7$ \citep{Looser+2023}. For these latter systems, the
sSFR decreased only in the last few tens (not hundreds) of Myr, 
meaning they are not `quiescent' according to the empirical 
definition, and may simply be undergoing temporary 
quenching \citep[e.g.,][]{Ceverino2018MNRAS.480.4842C,
dome_miniquenching_2023}.
For lower-mass quiescent systems, the uncertain strength
of internal feedback mechanisms compounds the effects of environment. Simulations show that the fraction of low-mass
quiescent satellites ($\Mstar = 10^9\text{--}10^{9.5}$~\MSun) should be as high as 0.3--0.4
already at $z=2\text{--}3$ \citep{donnari_activity_2019, donnari_quenched_2021}.
At the same time, as we have seen, galaxy overdensities appear to be already present at 
$z=5\text{--}7$, which suggests environmental effects may also start in earnest at these
epochs.

In this context, studying the properties of quiescent galaxies is critical to our
quantitative understanding of star formation and feedback models. Unfortunately, even with \jwst, our
picture of high-redshift quiescent galaxies is shaped by our ability to identify them;
this is particularly relevant at low stellar masses, where current spectroscopic observations
must rely on gravitational lensing \citep{Marchesini2023ApJ...942L..25M}, extremely young
systems \citep{Looser+2023}, or both \citep{Strait2023}. In this paper, we draw from the
unprecedented depth of our \jwst Advanced Deep Extragalactic Survey
\citep[JADES;][]{eisenstein_overview_2023, rieke_jades_2023,bunker_initial_2023} to extend 
the mass range of spectroscopically confirmed quiescent galaxies. We present \targetidfull,
a low-mass, quiescent system at $z=\zfinal$.
With a stellar mass $\Mstar = \mfinal$~\MSun and a quenching redshift $z=\zquench$,
this galaxy shows that feedback -- internal and/or environment-driven -- was in 
place and already capable of long-term quenching 2~Gyr after the Big Bang.
This galaxy is found only 35~pkpc away from a more massive central galaxy and has
an inverse colour gradient with radius, suggesting that
environment played an important role in sealing the fate of our target. 

In this work, we adopt the \citet{planck+2020} cosmology (their table~2),
a \citet{chabrier2003} initial mass function and the Solar metallicity of
\citet{Asplund2009}. All magnitudes are in the AB system \citep{oke+gunn1983},
all stellar masses are total mass formed and, unless otherwise specified, all
distances are proper distances.

\section{JADES Data}\label{s.jadesdata}

Photometry and spectroscopy for \targetidfull were obtained as part of our survey
JADES \citep{eisenstein_overview_2023}, a result of the joint \jwst/NIRCam \&
NIRSpec GTO teams. The data we use in this work has recently been publicly realeased
\citep{bunker_initial_2023, rieke_jades_2023}, and consists of NIRCam short- and long-wavelengths (SW, LW) photometry
and NIRSpec micro-shutter assembly (MSA) spectroscopy (PID 1180, PI: D. Eisenstein;
PID 1210, PI: N. L\"utzgendorf) in the GOODS-S field \citep{Giavalisco2004}. We also use 
medium-band photometry from the \jwst Extragalactic Medium-band Survey \href{https://dx.doi.org/10.17909/fsc4-dt61}{JEMS}
\citetext{PID 1963, PIs: C.~C.~Williams, S.~Tacchella \& M.~Maseda; \citealp{williams_jems_2023}}
and archival photometry from both \hst/ACS \citep[from GOODS;][]{Giavalisco2004} and 
\hst/WFC3 IR \citep[from CANDELS;][]{grogin+2011}. For \hst, 
we use data re-processed as described in \citet{illingworth+2016} and \citet{whitaker+2019}.

For the MSA spectroscopy, we use the prism/clear observations, 
spanning wavelengths 0.6--5.3~\mum with a resolution $R=30\text{--}300$.
Even though medium-resolution spectroscopy is also available 
\citep{bunker_initial_2023}, in this paper we focus only on the
low-resolution data because the accompanying grating 
spectra have insufficient signal-to-noise ratio.
The observations used a 3-shutter slit
with nodding for background subtraction and dithering to sample different detector regions.
The target allocation is described in \citet{bunker_initial_2023}, and was optimised using
the \textsc{eMPT} software \citep{bonaventura_optimal_2023}.
Our galaxy was originally prioritised as a \lya-dropout candidate at $z=9$ and had a high 
priority; as a result, its prism/clear exposure time was 28~hours.
We will discuss later in the paper why we now believe this high-redshift
candidate is actually a lower-redshift ($z=\zfinal$) galaxy with an
evolved stellar population. 

The data was reduced using software by the ESA NIRSpec Science Operations Team (SOT) and
by the NIRSpec GTO Team. The detailed procedure is described in 
\citet{Curtis_Lake2022arXiv221204568C, bunker_initial_2023, Carniani23, Curti2023MNRAS.518..425C} and Carniani et~al. (in~prep.). We note that
the spectrum was corrected for wavelength-dependant path losses assuming a point-source 
distribution. This is only a first-order approximation, but in the analysis we upscale the
spectrum to match the photometry.

\section{Photometry}\label{s.photometry}

\begin{figure*}
   \centering
   \includegraphics[width=\textwidth]{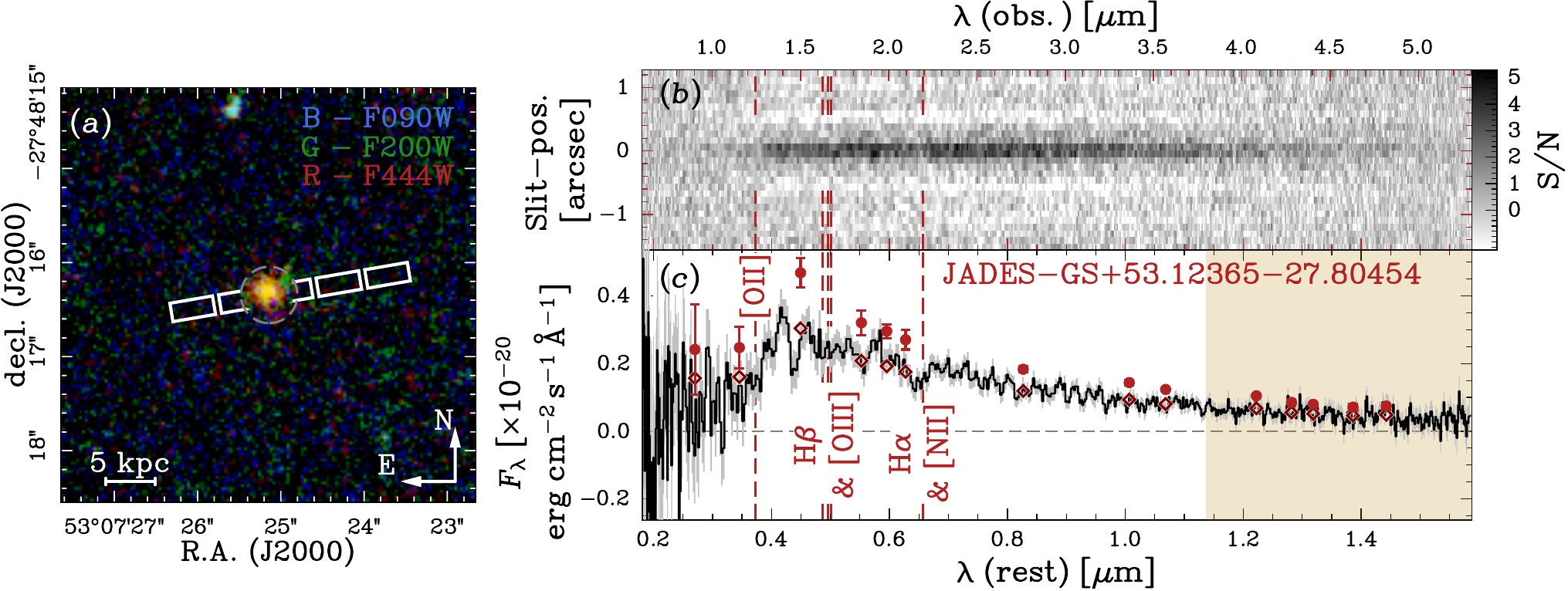}
   {\phantomsubcaption\label{f.img.a}
    \phantomsubcaption\label{f.img.b}
    \phantomsubcaption\label{f.img.c}}
   \caption{NIRCam false-color image (panel~\subref{f.img.a}), using F444W/F200W/F090W
   in the red/green/blue channels. The galaxy displays a compact morphology, with possible
   signs of a neighbour/merger signatures to the north-west. The rectangles show the nominal
   position of the NIRSpec/MSA micro-shutters 
   (for clarity, the two shutters overlapping the galaxy are cropped) and the
   dashed grey circle is the 0.3-arcsec radius aperture we used for the photometry.
   The 2-d spectrum (panel~\subref{f.img.b}) shows a spectral break at 1.25~\mum, and
   tentative evidence for flux blueward of the break.
   The 3-pixel boxcar extraction (panel~\subref{f.img.c}) also shows
   flux detection blueward of the break. At the same time,
   we do not detect any strong nebular emission at the wavelengths corresponding to the
   \lya-drop solution (shaded region to the right) -- supporting the hypothesis that this 
   galaxy is at $z=\zfinal$. Even at this fiducial redshift, no strong nebular emission is 
   detected (dashed red vertical lines). The red circles with errorbars are the \jwst/NIRCam
   circular photometry, which is $1.5\times$ brighter than the aperture-corrected spectrum; for
   comparison, we show the photometry downscaled by a factor of 1.54 as small diamonds.
   The strong dips in the spectrum at 1.5 and 2.2~\mum are
   likely to be outliers, perhaps due to correlated noise; bootstrapping the individual
   nods and summing, both these features disappear, which suggests they are
   not real.}
   \label{f.img}
\end{figure*}

The target imaging and spectrum are presented in Fig.~\ref{f.img}. Panel~\subref{f.img.a}
shows a false-colour RGB image (using \jwst/NIRCam F444W, F200W and F090W). \targetidfull 
appears compact in size and green/yellow in colour, underscoring the dearth of light in the 
blue channel. Using \forcepho, we model the light distribution as a
S\'ersic function \citep{sersic1968}, and obtain a half-light semi-major axis $\re=0.72\pm0.02$~kpc\footnote{Here and in the following, we always convert
apparent sizes and magnitudes to physical values assuming $z=2.34$, following
the determination in \S~\ref{s.redshift}.}
an index $n=1.00\pm0.07$ and an axis ratio of $0.91\pm0.04$, suggesting the
galaxy is a disc seen nearly face on (\S~\ref{a.photometry}).

There is a relatively extended feature to the north west, a possible 
sign of recent interaction. This feature appears most prominently in F200W,
which would include \Halpha at $z=2.34$.
However, no \Halpha is visible in the NIRSpec data (Fig.~\ref{f.img.c}).
The extended feature is not clearly visible in any other filter, which 
suggests a nebular origin. If this interpretation was correct, we would 
then expect an \oiii counterpart to \Halpha, which would fall within the
range of the F150W filter. However, the F150W flux of this extended 
region is clearly much lower than F200W (see Fig.~\ref{f.forcepho}),
suggesting that no \oiii is 
present. The absence of the \oiii counterpart in F150W 
could be due to the fact that, at this redshift, the rest-frame 
0.5008-\mum line of \oiii falls very near the sensitivity gap between 
F150W and F200W. Further, deeper observations may be needed to
understand the physical nature of this feature.

The galaxy spectrum (panels~\subref{f.img.b} and~\subref{f.img.c}) appears
systematically fainter than the \jwst photometry (red circles with errorbars in
panel~\subref{f.img.c}). This shows that (for this target) the point-source 
aperture correction is insufficient, and an additional upscaling is needed. The 
small red diamonds are the observed photometry downscaled by a factor 1.54;
the good agreement between the spectrum and the uniformly scaled photometry 
suggests that the missing aperture correction is approximately achromatic.
The value of 1.54 was determined using spectral energy distribution modelling (see 
\S~\ref{s.sed}).

\begin{figure}
   \centering
   \includegraphics[width=\columnwidth]{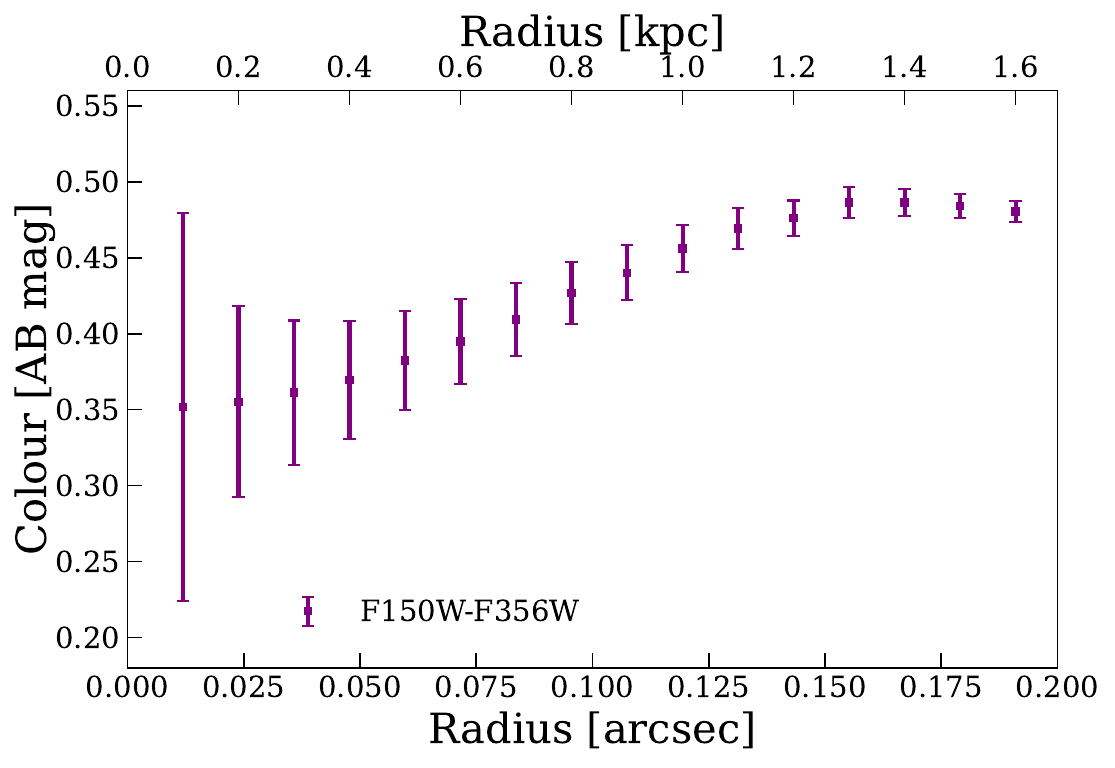}
   \caption{Radial colour gradient of our target, from PSF-matched
   NIRCam photometry. The centre is bluer than the outskirts, i.e., this galaxy
   displays an `outside-in' colour structure.
   }\label{f.colour}
\end{figure}

In Fig.~\ref{f.colour} we show the F150W-F356W radial colour gradient. This was
obtained by smoothing both the F150W and F356W images to the point-spread function (PSF) at 
4.4~\mum, and then measuring the flux inside concentric
elliptical annulii, following the shape and position angle of the \forcepho model
\citep{baker_insideout_2023}. The galaxy displays an `outside-in' colour gradient,
with the centre bluer than the outskirts.

\section{Redshift determination}\label{s.redshift}

\begin{figure}
   \centering
   \includegraphics[width=\columnwidth]{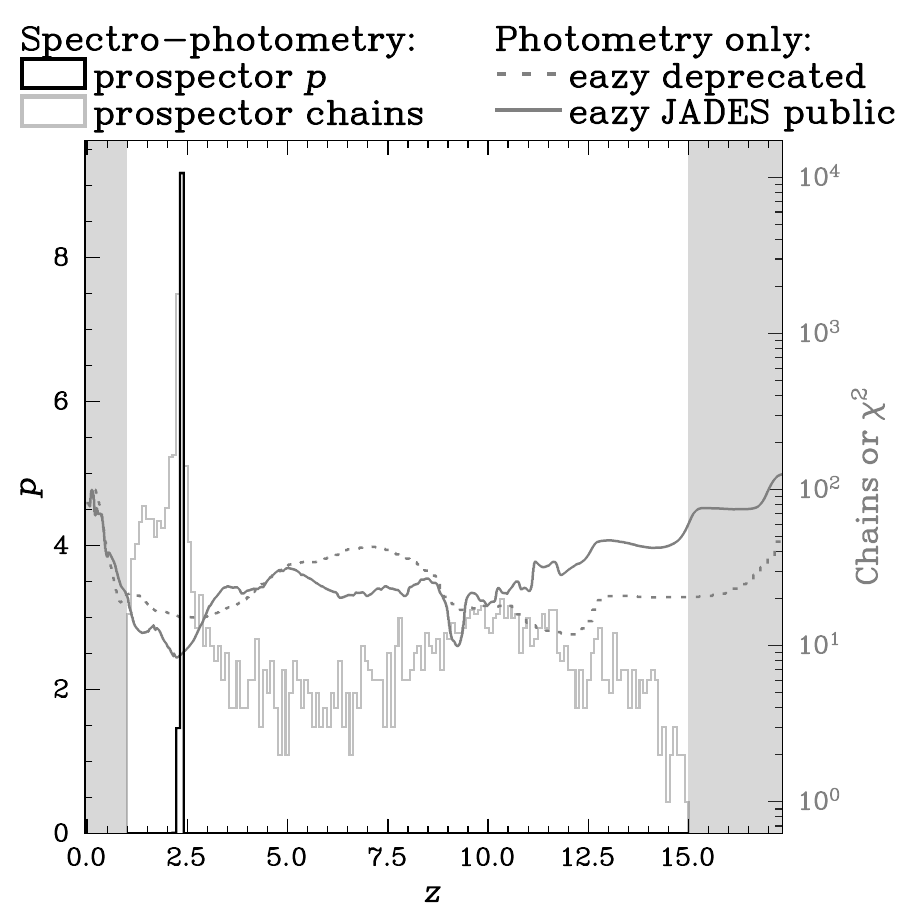}
   \caption{Distribution of redshifts for our target, based on the \eazy
   $\chi^2$ distribution (solid grey line) and on the \prospector posterior
   distribution (solid black histogram). The dashed grey line shows the
   \eazy results from the initial, preliminary photometry; the minimum 
   $\chi^2$ at $z_\mathrm{phot}=11.5\text{--}12$ explains why this source 
   was included in JADES. The updated photometry shows excellent agreement with the \prospector posterior distribution. The grey histogram shows the chains from the \prospector optimisation, demonstrating that the 
   software explores the high-redshift parameter space.
   }\label{f.redshift}
\end{figure}

The spectrum shows a clear break at 1.25~\mum. The photometric 
redshift from the publicly available JADES catalogue \citep{rieke_jades_2023} 
is $z_\mathrm{phot} = 2.18$, with a range $\sigma_\mathrm{68,low}\text{--}
\sigma_\mathrm{68,high}$ of 1.96--2.79; these values were measured from \hst and
\jwst photometry using the \eazy 
template-fitting software \citep{brammer_eazy_2008}; $z_\mathrm{phot}$ 
corresponds to the minimum \textchi$^2$ solution ($z_\mathrm{a}$ in \eazy) and 
$\sigma_\mathrm{68,low}\text{--}\sigma_\mathrm{68,high}$ is the 68-per-cent confidence interval 
from the redshift probability distribution \citep{hainline_cosmos_infancy_2023}.
The $\chi^2$ curve is reported in Fig.~\ref{f.redshift} (solid grey 
line) and shows an additional, secondary minimum at $z \approx 9$. The 
$\chi^2$ photometry-only curve originally used for the target selection is also reported (dashed 
grey line); this curve used one of the first photometric data releases (internal release v0.4), based 
on a preliminary reduction of the NIRCam data which is now deprecated. The
resulting $\chi^2$ curve showed a primary minimum at $z_\mathrm{phot}
\approx 11.5\text{--}12$, which explains the inclusion of this galaxy as a 
high-redshift candidate in JADES.

We go beyond simple photometric SED fitting, by taking into account the information provided by the NIRSpec spectrum (see \S~\ref{s.sed} for the model description).
If we run \prospector with an initial redshift guess of $z=7$ and a
uniform redshift prior between $1<z<15$, the resulting redshift posterior is $z=2.34^{+0.01}_{-0.06}$, which rules out the high-redshift solution at very high confidence (black
histogram in Fig.~\ref{f.redshift}). We also show the distribution of the
chains from the optimisation algorithm (grey histogram), illustrating that
\prospector does explore high-redshift solutions, but finds them very unlikely
\textit{vis-{\`a}-vis} the data. However, we note that this model does not
properly take into account the shape of the damping wing, therefore we consider
it  only marginal evidence.

We therefore consider an initial redshift estimate of $z=2.34$, from interpreting 
the break as a Balmer break. In addition to the \eazy result, this
redshift is based on three lines of evidence. First, the break has a 
smooth shape suggesting a Balmer/4000-\AA break instead of the 
much sharper break expected from a \lya drop. Secondly, for a $z=9$ object, the spectrum lacks 
both emission lines (ruling out a star-forming solution) and a Balmer 
break (ruling out a quenched stellar population; all these features would 
fall inside the shaded region in panel~\subref{f.img.c}). Finally, the object is tentatively detected
(3--4~\textsigma) blueward of the putative \lya drop at 1.2~\mum
(panel~\subref{f.img.b}); to further confirm, we also integrate the NIRSpec flux between 0.9~\mum and 1.1~\mum. The value of
1.1~\mum also accounts for a 0.1~\mum (27,000~\kms) buffer to exclude a
proximity zone which may otherwise smooth the \lya drop
\citep[e.g.,][]{Curtis_Lake2022arXiv221204568C}.
Using an unweighted sum and the nominal uncertainties, we obtain a 3.8-\textsigma detection;
bootstrapping the data one hundred times (by random sampling with replacement) and estimating
the uncertainties from the resulting distribution of fluxes, we obtain a 4.7-\textsigma
detection.
These various evidences strongly favour the solution at $z=2.34$ over the one at $z=9$ (see \S~\ref{ss.z9} for a further discussion).

The spectral shape redward of the break favours the Balmer break over the
4000-\AA break, but, in the following analysis, we always provide a redshift range
sufficiently wide to include the lower-redshift solution ($z\approx2.1$) corresponding to
the 4000-\AA interpretation. Measuring the $F_\lambda$ break strength based on the
rest-frame wavelength ranges used by \citet{Curtis_Lake2022arXiv221204568C},
we obtain a value of $2.7\pm0.7$, which is statistically consistent
with (but somewhat higher than) the maximum Balmer-break strength 
allowed by models \citetext{\citealp{Curtis_Lake2022arXiv221204568C}
find a maximum value of 2; if we use simple stellar populations with 
the C3K model atmospheres \citealp{conroy+2019} and MIST isochrones 
\citealp{choi+2016}, we find a maximum Balmer-break value of 2.2 for a 
stellar population with metallicity $\log Z/\ZSun = -2$ and age 
0.7~Gyr}.

\section{SED modelling}\label{s.sed}

To marginalise over model assumptions and implementation, we use two 
different approaches to model the target SED. We use the Bayesian software packages \beagle
\citep{chevallard_beagle_2016} and \prospector \citep{Johnson2021ApJS..254...22J}.

\begin{figure*}
   \centering
   \includegraphics[width=\textwidth]{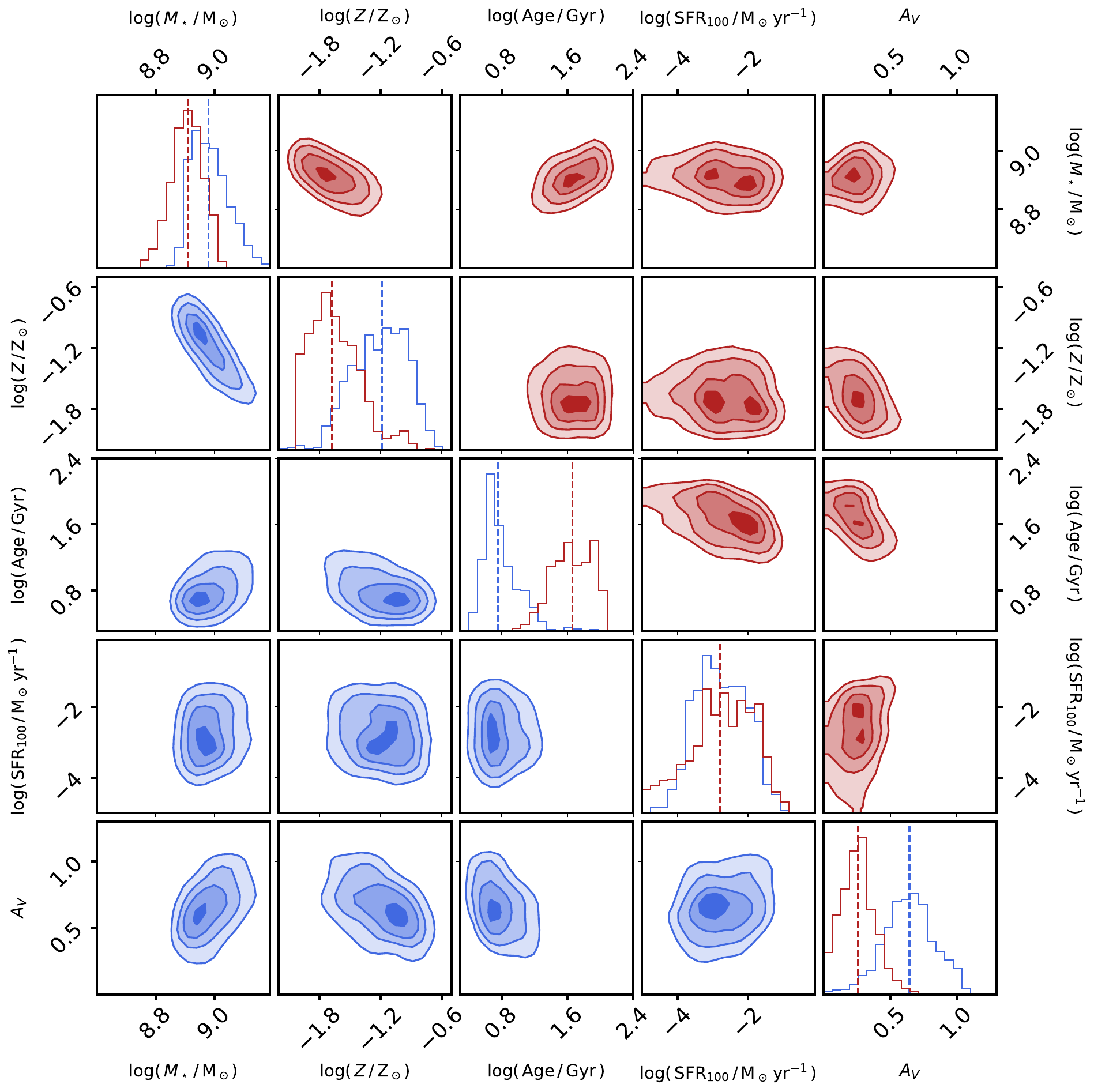}
   \caption{Marginalised posterior distributions showing key physical parameters from
   our SED modelling using \beagle (blue) and \prospector (red). The \beagle $M_\star$ and $\mathrm{SFR}_\mathrm{100}$ posteriors include the 1.54 aperture correction factor derived from \prospector. The two models
   agree remarkably well, except for the (notably degenerate) age, metallicity and dust
   attenuation. Part of these differences are due to the inclusion of photometry in the
   model inference with \prospector, but even without photometry, some difference remains.
   We interpret these remaining differences as arising from the different SFH
   parametrisations, which favour younger/older ages for \beagle/\prospector. To break
   the degeneracy and determine which solution is most accurate we would require higher 
   resolution spectroscopy and/or longer wavelength coverage/sensitivity.}\label{f.corners}
\end{figure*}

We use \beagle to fit to the prism/clear spectrum (we assumed a flat cosmology with $H_0 = 
70$~\kms~Mpc$^{-1}$ and $\Omega_\mathrm{m}=0.3$, but corrected the resulting masses and
SFRs by five per~cent to match the default cosmology). Our model assumes an initial delayed 
exponential SFH 
with free parameters maximum stellar age, $t / \mathrm{yr}$, and location of the peak of star formation, $\tau / \mathrm{yr}$. The most recent star formation is modelled as a constant SFH which allows for additional flexibility, with free parameters $\mathrm{SFR}_\mathrm{const} / \MSun\peryr$ and duration, $t_{\mathrm{const}} / \mathrm{yr}$ (this recent burst is then
assigned a negligible SFR, hence the quiescent interpretation).
The nebular emission is characterised by three ionised gas parameters: the interstellar metallicity, $Z_{\mathrm{ISM}}/\ZSun$, the ionization parameter, \Us, and the mass fraction of interstellar metals locked within dust grains, \xid. Dust attenuation is modelled with two components (ISM attenuation which is applied to all stars, plus an additional birth cloud attenuation which is only applied to stars younger than 10~Myr) following the prescription of \cite{Charlot2000ApJ...539..718C}. We fit for the total effective \textit{V}-band attenuation optical depth, $\tauV$, and fix the ratio of the \textit{V}-band ISM attenuation to the \textit{V}-band ISM + birth cloud attenuation to $\mud=0.4$. We also fit for redshift, $z$, stellar metallicity, $Z/\ZSun$, and total stellar mass, $\Mstar/\MSun$. In total the
\beagle model has 11 free parameters, of which their prior distributions are shown in
Table~\ref{tab.priors} (left two columns). Note that, in the following, we upscale the
\beagle-inferred \Mstar and SFR by an aperture correction of 1.54 (0.2~dex), determined using 
\prospector.

\input{model_table}

\input{posterior_table}

For \prospector, we model jointly the spectro-photometric SED, including both the NIRSpec
data as well as broad- and medium-band \jwst/NIRCam photometry \citetext{from JADES and JEMS}
and archival \hst/ACS and WFC3 data.
This setup enables us to simultaneously measure the detailed spectral features, extend the
wavelength range, and capture a photometric aperture correction. To capture the varying
spectral resolution of the prism spectrum, we use the prism nominal line-spread function
\citep{jakobsen+2022}. The effective line-spread function depends on the size of the
target relative to the width of the MSA shutter, but including this correction
\citetext{de~Graaff et~al., in~prep.; see e.g., \citealp{maiolino_a_small_2023} for
a description} does not substantially alter our conclusions.

Our methodology follows the approach of \citet{tacchella_fast_2022, tacchella_early_2022}.
We use eight
log-spaced age bins from the time of observation to the age of the
Universe. The last time bin is narrower than the others, but we verified that removing it 
from the fit (i.e., assuming SFR=0 in that bin) does not change our conclusions, except for
an older mass-weighted age.
The SFR is parametrised as the log ratio between adjacent time bins, giving seven
free parameters (one parameter is captured by the total stellar mass formed).
These ratios follow a Student's t prior with mean 0, standard deviation 0.3 and
$\nu=2$, which prefers continuous SFHs \citep{leja+2019a}.
We use a dust model with differential dust attenuation towards the birth
clouds \citep{Calzetti1994, Charlot2000ApJ...539..718C}. The diffuse dust component is
parametrised by the optical depth at 5,500~\AA, \tvdiff, and by the dust index $n$, which
is tied to the strength of the UV bump \citep{kriek_conroy_dust_2013}.
The attenuation towards the birth clouds is equal to the diffuse attenuation above, plus an
additional dust screen as in
\citet{Charlot2000ApJ...539..718C}, parametrised by the optical depth $\tvbc = f_\mathrm{d} \cdot \tvdiff$. The free parameters and prior distributions for \prospector are summarised in Table~\ref{tab.priors} (right two columns).

\begin{figure*}
   \centering
   \includegraphics[width=\textwidth]{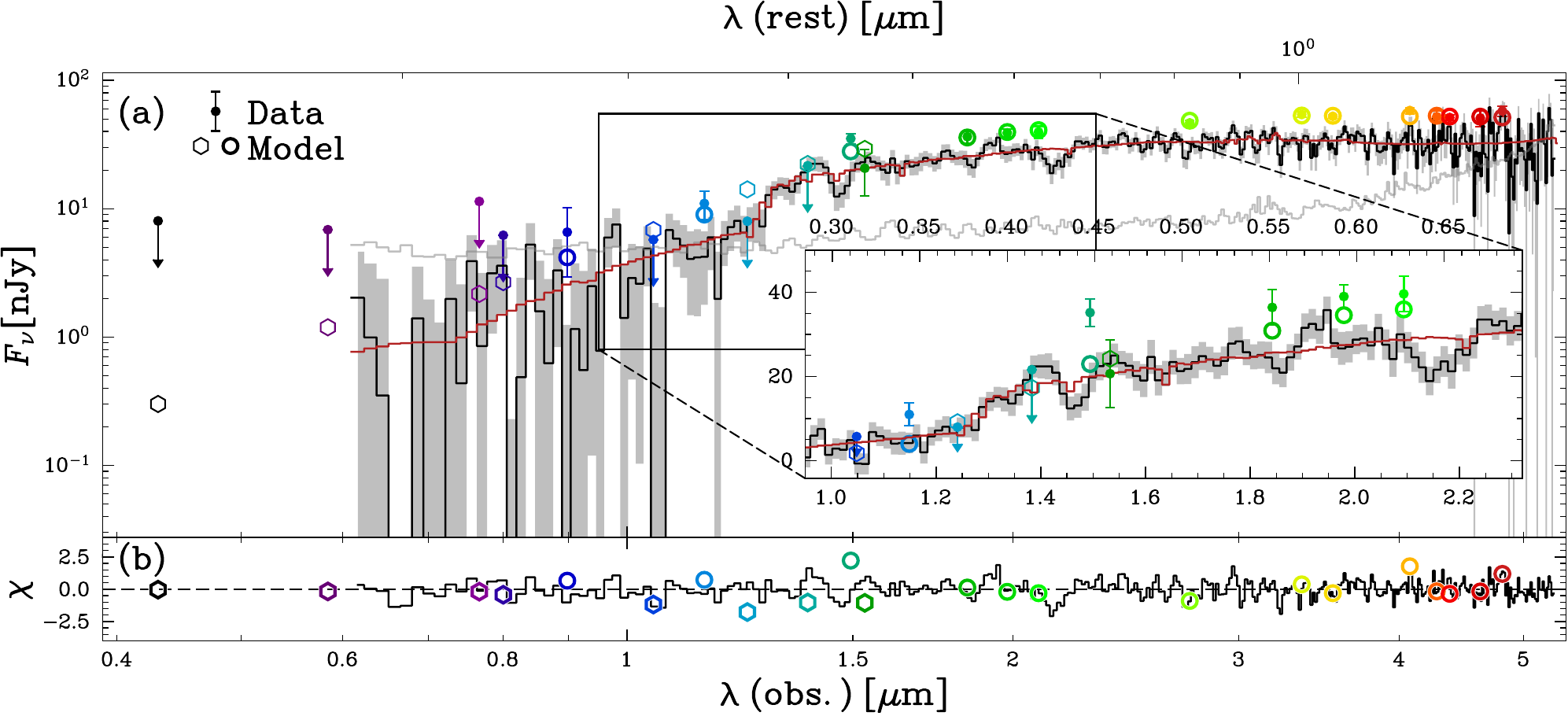}
   {\phantomsubcaption\label{f.prosp.model.a}
    \phantomsubcaption\label{f.prosp.model.b}}
   \caption{Summary of the data and \prospector model. In panel~\subref{f.prosp.model.a}, 
   the black line and the grey shaded region are the NIRSpec spectrum and uncertainties, the red solid line is the maximum a-posteriori \prospector model,
   and the thin grey line is the 2-\textsigma uncertainty on the NIRSpec data.
   The small circles with errorbars are the photometric data from \hst/ACS, \hst/WFC3 IR and \jwst/NIRCAM; the arrows are 1-\textsigma upper limits; the larger thin hexagons and thick circles are synthetic photometric measurements from \prospector for \hst and \jwst, respectively.
   \prospector infers a 1.54 scaling factor between spectrum and photometry. The bottom panel shows the residuals divided by the uncertainties.
   }\label{f.prosp.model}
\end{figure*}

\begin{figure*}
   \centering
   \includegraphics[width=0.8\textwidth]{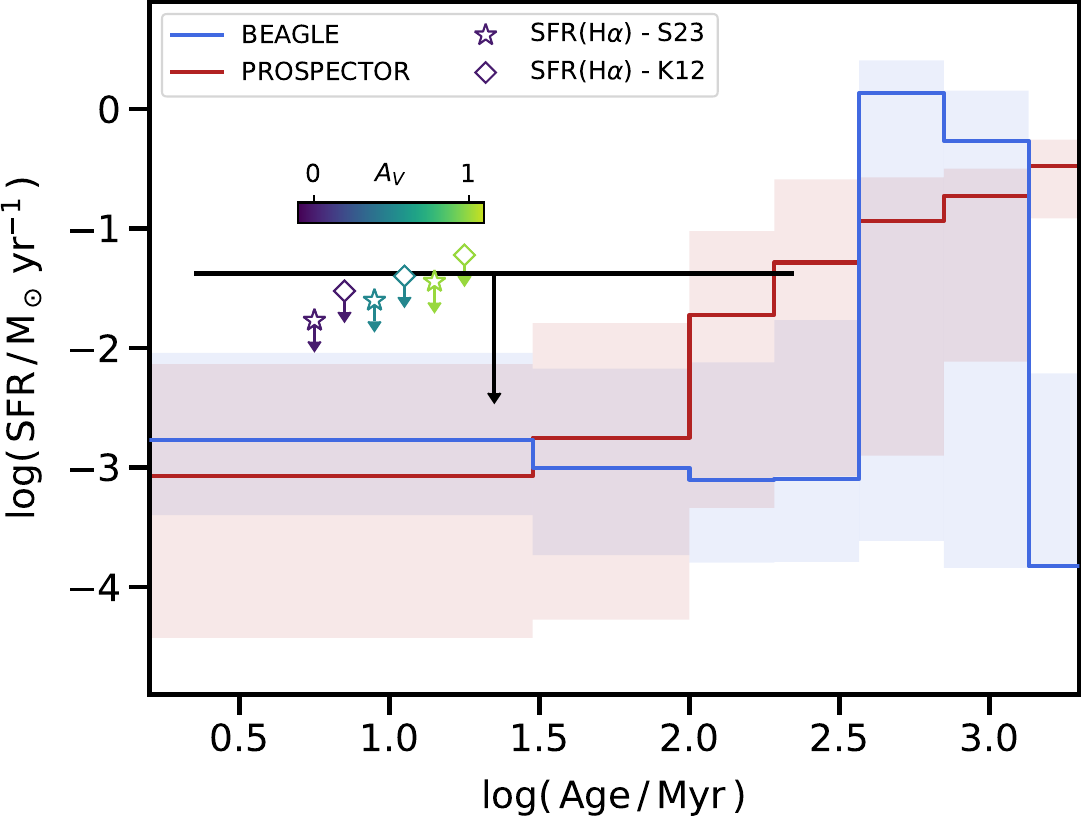}
   \caption{Parametric SFH from \beagle (blue, shaded area encloses the 
   16\textsuperscript{th}--84\textsuperscript{th} percentiles) and \prospector (red).
   The \beagle analytical SFH has been rebinned onto the \prospector grid for ease of
   comparison.
   Both models agree that the galaxy has been quiescent for
   hundreds of Myr. The black horizontal upper limit is the quiescent threshold at
   $z=2.34$ \citep[e.g.,][]{pacifici+2016}. Stars/diamonds are 3-\textsigma upper limits
   on the \Halpha-derived SFR, using the \citet[][S23]{shapley2023}/\citet[][K12]{kennicutt_star_2012} calibrations and increasing values
   of the nebular attenuation $A_{V,\mathrm{neb.}}$, as indicated by the symbol colour
   (these upper limits all correspond to an age of 10~Myr, and are offset horizontally
   for display purposes).
   }\label{f.sfh}
\end{figure*}

\begin{figure*}
   \centering
   \includegraphics[width=0.9\textwidth]{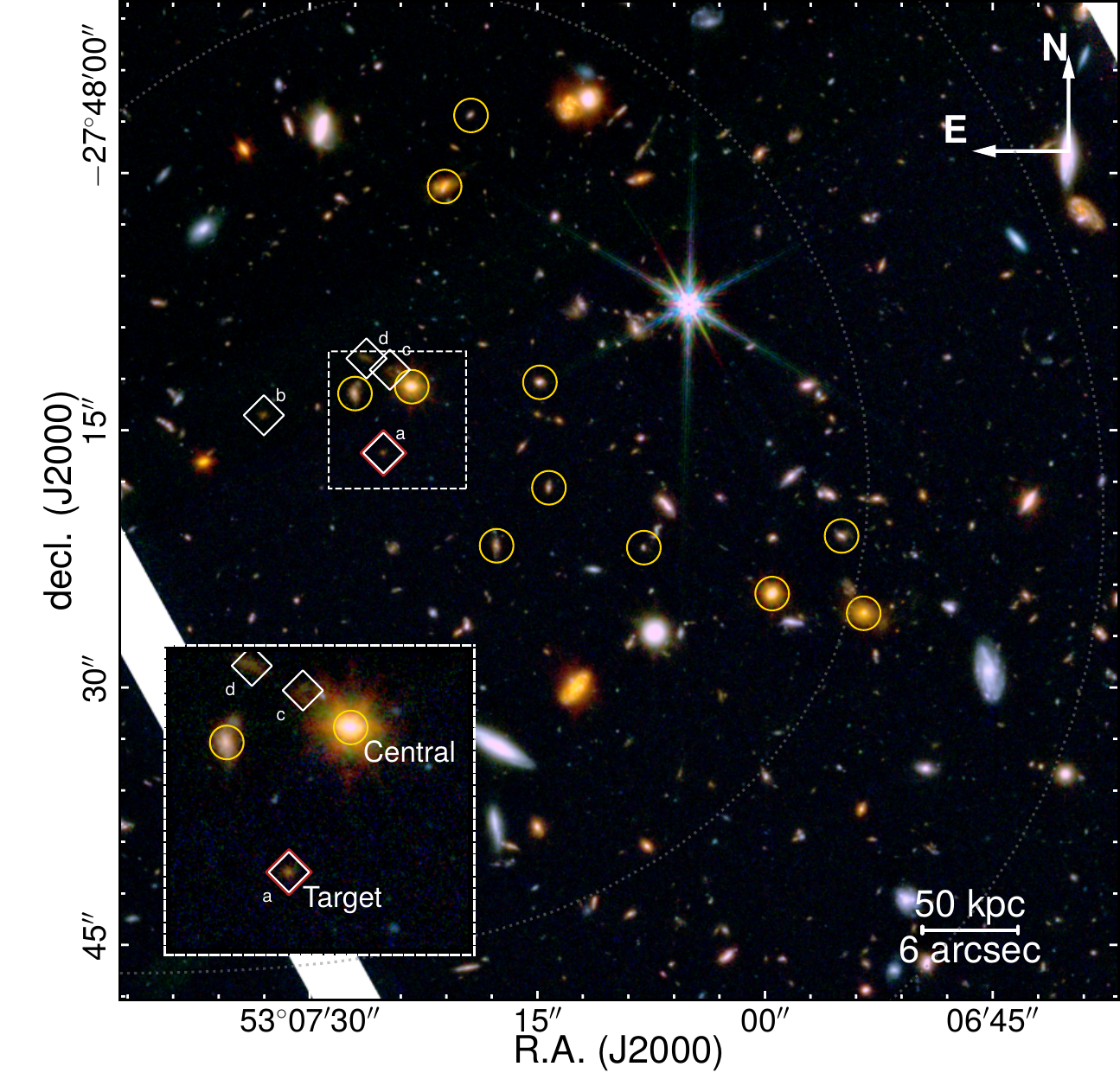}
   \caption{The environment surrounding our low-mass quiescent galaxy (red diamond), centred on the light-weighted centre of the structure (halfway
   between the most massive galaxy and the most massive pair). The inset shows
   our target and the most massive galaxy. Candidate group members are
   highlighted by the yellow circles; these are selected from robust
   photometric redshifts or spectroscopic redshifts. The smaller, white diamonds are a selection of fainter and
   close ($<10$~arcsec) photometrically selected Balmer break galaxies, showing possible
   evidence for additional quenched satellites. The grey dotted curves are the contours
   of 3- and 4-\textsigma significance for the photometric overdensity (cf.~Fig.~\ref{f.overdensity}).
   }\label{f.group}
\end{figure*}

\begin{figure*}
   \centering
   \includegraphics[width=0.9\textwidth]{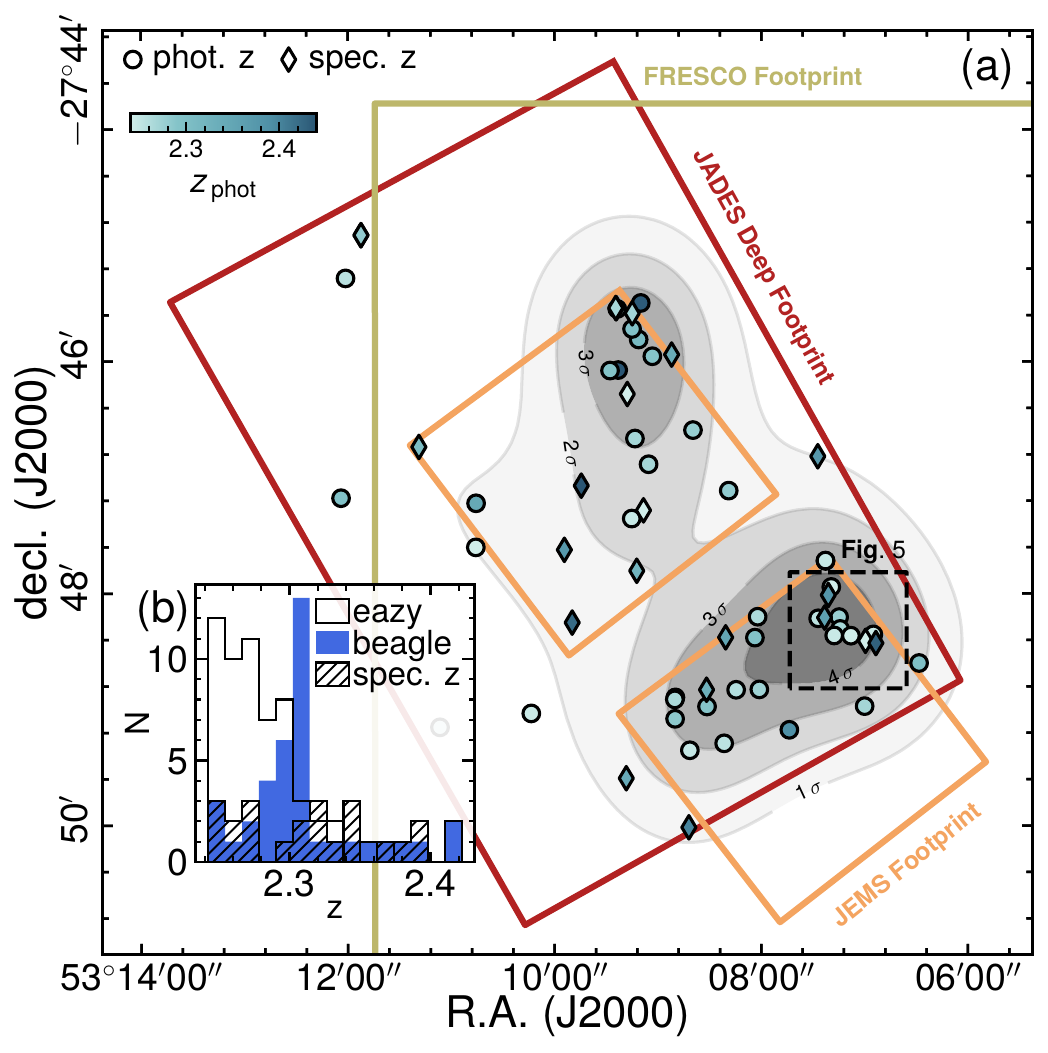}
   {\phantomsubcaption\label{f.overdensity.a}
    \phantomsubcaption\label{f.overdensity.b}}
   \caption{Overdensity surrounding our target; circles/diamonds represent individual 
   galaxies colour coded by their robust photometric redshifts/spectroscopic redshifts.
   All galaxies have redshifts in the range $2.24<z<2.44$ and lie in projection within 
   10~cMpc from the central galaxy \targetidcentral.
   The grey contours trace lines of constant overdensity significance above the background
   noise at this redshift (1, 2, 3 and 4-\textsigma contours), estimated using a KDE with
   smoothing scale 1~cMpc. The dashed black square is the region displayed in Fig.~\ref{f.group}. The bottom-left inset 
   (panel~\subref{f.overdensity.b}) shows a histogram of the \eazy 
   photometric redshifts (empty black histogram), \beagle
   photometric redshifts (filled blue), and spectroscopic redshifts
   (hatched histogram).
   Also shown are the footprints of the
   main surveys used for this measurement. The peak of the overdensity lies only
   3.5~arcsec away from the our target and 5.4~arcsec away from the central galaxy.
   }\label{f.overdensity}
\end{figure*}

\section{A low-mass quiescent galaxy at redshift 2.34}\label{s.results}

In the following, we focus on the results from \beagle, but we stress that the reported
values are consistent with the results of the \prospector analysis (provided we upscale the
\beagle extensive quantities like \Mstar and SFR by a factor 1.54 to account for the additional
calibration offset between the spectrum and photometry, see below). When they are present, we
highlight the differences, which seem invariably due to different data
(spectro-photometry vs photometry only) and different SFH priors between
the two models. Our interpretation would be unchanged using everywhere the
\prospector results. Table~\ref{tab.posteriors} shows the \beagle and \prospector posterior medians for parameters of interest and their corresponding 16th and 84th percentiles. \beagle SED modelling finds a redshift $z=\zfinal^{+0.02}_{-0.02}$ and a stellar mass of $\log(\Mstar/\MSun) = \logmfinal^{+0.07}_{-0.06}$. 
The current SFR (averaged over the last 100~Myr) is consistent with no star
formation: $\log(\,\mathrm{SFR}_\mathrm{100} \,/\, \mathrm{M}_\odot \, \mathrm{yr}^{-1}\,) = -2.8^{+0.8}_{-0.7}$; placing \targetidfull firmly in the realm of 
low-mass, quiescent galaxies (Fig.~\ref{f.corners}, blue histograms). \prospector confirms
these results: $\log(\Mstar/\MSun) = 8.91^{+0.05}_{-0.06}$ and
$\log(\,\mathrm{SFR}_\mathrm{100} \,/\, \mathrm{M}_\odot \, \mathrm{yr}^{-1}\,) = -2.8^{+1.0}_{-1.5}$ (notwithstanding the differences in 
implementation, model assumption, priors, and the simultaneous use of the
spectrum in addition to photometry; see Fig.~\ref{f.corners}, red 
histograms). In particular, \prospector infers
a median upscaling between the spectrum and photometry by a factor of 1.54 
(or 0.2~dex; Fig.~\ref{f.prosp.model}); without this upscaling, the 
\beagle-inferred median mass was $\log(\Mstar/\MSun) = 8.79$, still within 25~per cent from the \prospector result.

Even though both models find low stellar metallicity and dust attenuation, the two distributions are
not consistent, with \prospector finding lower metallicity and dust attenuation than
\beagle; this may be connected to the difference in mass-weighted stellar ages. \prospector
finds an older overall age than \beagle.
A key difference between \beagle and \prospector is the inclusion of photometric data to
constrain the \prospector model. If we repeat the \prospector model inference ignoring photometry, 
we find indeed a younger median age and higher median dust attenuation; even though these values
are formally consistent with the default, spectro-photometric inference, they are closer to the
\beagle results. The remaining difference is most likely due to the
different SFH parametrisations and priors: we expect the earliest phases of the SFH to be
poorly constrained by the data. Looking at the SFHs of the two models (Fig.~\ref{f.sfh}), there is a large difference in SFR at the earliest time bin. The large SFR in \prospector 
arises from the high SFR in the adjacent time bin combined with the continuity prior, whereas the low 
SFR in \beagle arises from the well constrained SFH peak and from the shape of the delayed 
exponential SFH. Both SFHs peak between 0.5--1~Gyr prior to observation, then 
decline to low SFR. The decline is more rapid in \beagle than \prospector, as expected from 
the continuity prior we used in \prospector, which biases the SFH against sudden changes in SFR.

Quiescence is empirically confirmed by the rest-frame UVJ colours, with
$V-J=0.28\pm0.05$~mag and $U-V=1.44\pm0.11$~mag, \targetidfull satisfies the
conditions for quiescence at this redshift \citep[e.g.,][]{williams+2009, Ji+2023}.
Further, independent confirmation comes from the lack of nebular emission; with the depth of JADES, we can
place stringent upper limits on the short-timescale SFR. We obtain a 3-\textsigma upper limit $F(\Halpha)
< 1.8\times10^{-19}$~\fluxcgs. The conversion between this value and the SFR
is quite uncertain, lacking any direct constraint on the gas metallicity and
attenuation, therefore we present a range of possibilities. Using the
low-metallicity SFR calibration of \citet{shapley2023}, the upper limit on $F(\Halpha)$ translates into a SFR upper limit of
SFR$<0.02$~\MSun~\peryr (with no dust attenuation; dark purple star in Fig.~\ref{f.sfh}; all \Halpha-derived
SFR upper limits are placed around $t=10$~Myr and offset along the time axis for display purposes).
Considering $A_V=0.5$ or 1~mag \citetext{and using the \citealp{cardelli+1989} extinction curve with
$R_V=3.1$}, we find upper limits of 0.03 and 0.04~\MSun~\peryr, respectively (blue and green stars).
Switching to the solar-metallicity calibration of \citet{kennicutt_star_2012} these upper limits increase by
50~per cent (diamonds).

\section{Surrounding environment}\label{s.env}

We searched for neighbours in the catalogues of
\citet{salimbeni_comprehensive_2009} and \citet{chartab_largescale_2020}. The
former contains three overdensities at $z=2.3$, but they are too far in
projection from our galaxy to be physically associated with it 
(1.5\textdegree). In the second catalogue, we find
nineteen galaxies within a radius of 120~arcsec (1~proper Mpc) and within $\pm2,500$~\kms 
from $z=\zfinal$. Of these objects, one lies particularly close: galaxy \targetidcentral is 
only 4.2~arcsec to the north west of our target ($\approx$35~kpc in projection) and has a 
spectroscopic redshift $z=2.3490$ \citep{wuyts_optical_2009}. For this
galaxy (labelled `Central' in Fig.~\ref{f.group}) we determine a 
tentative mass of $\Mstar=1.7 \times10^{11}$~\MSun with little SFR 
(within the last 100~Myr, $\mathrm{SFR}=1.8$~\MSun~\peryr; see Appendix~
\ref{a.central}). This galaxy is therefore more than two orders of 
magnitude more massive than our main target \targetidfull.
From here on, we refer to this galaxy as the central galaxy in the structure; this is based
on both its high stellar mass and its location near the centre of an overdensity (see below).
The central galaxy is compact (half-light radius $\re=0.5\pm0.1$~kpc), has a high S\'ersic 
index \citep[$n=5.7\pm0.1$;][]{cassata_constraining_2013} and a low sSFR
($\mathrm{sSFR}=0.01$~Gyr$^{-1}$). Taken together,
these measurements suggest the central galaxy may be an evolved, quiescent system not unlike 
other massive quiescent galaxies at redshifts 1--3.
To the south west, 24.2~arcsec away from this central (0.66~cMpc), lies a pair of massive 
galaxies; these have spectroscopic redshifts of 2.252 and 2.208 determined from \hst grism 
spectroscopy \citep[and are consistent with $z=2.34$ given the large
uncertainties associated with grism spectroscopy of Balmer/4000-\AA
break galaxies;][]{momcheva_3dhst_2016}.
Together with the central, this pair are the three most massive galaxies in the overdensity 
($\Mstar = 1.1\times10^{11}$ and $8.7\times 10^{10}$~\MSun) and all three show a clear 
Balmer break (1~mag decrease) between F115W and F200W. We can 
rule out an extremely dusty SED thanks to the combination of JADES and JEMS photometry, and
find again negligible SFR ($<0.1$ and 2.6~\MSun~\peryr). These findings suggest that the most massive galaxies in this environment are already fairly evolved by $z=2.34$. 

Furthermore, we identify a clear galaxy overdensity at $2.24<z<2.44$ (within
0.1\footnote{Note that while this redshift tolerance is much higher
than the typical velocity dispersion of galaxy clusters (10,000~\kms),
we are limited by the precision of the photometric redshift
measurements.}
from the systemic spectroscopic redshift of our target) using all of the available photometry and spectroscopy 
\citetext{JADES, JEMS and FRESCO \citealp{oesch_fresco_2023}}. We 
searched within 10 comoving Mpc from the central galaxy (6~arcmin),
using PSF-matched Kron photometry and requiring magnitude brighter than 
28~mag in F200W, F277W, F356W and F444W. We require these galaxies to be 
bright in these four filters because these are the four broad-band 
filters that are stacked to create the photometric detection image. We do 
not make any requirements on shorter wavelength filters in order not to 
bias against SW dropouts (e.g. F090W, F115W, or F150W).
We use again the $z_\mathrm{phot}$ table from \citet{rieke_jades_2023}
to select galaxies in the redshift interval $2.24<z_\mathrm{phot}<2.44$.
However, we consider only well-constrained photometric redshifts, i.e., 
galaxies with $\mathrm{zConf}<0.5$, where $\mathrm{zConf}\equiv\sigma_\mathrm{68,high}-\sigma_\mathrm{68,low}$
and can be thought of as twice the standard deviation on $z_\mathrm{phot}$. We require photometric redshifts to be well constrained so that any
large-scale structures that we might identify
are robust, rather than chance projections resulting from poor photometric redshift solutions \citep{helton_jades_2023}.
These targets are complemented by objects with spectroscopic redshifts from publicly
available and independently reduced FRESCO data, based on detecting \Pabeta.
In practice, we search for 5-\textsigma emission-line detections in the
observed wavelength range $4.153 < \lambda < 4.410$~\mum, corresponding
to the range of \Pabeta at $2.24 < z < 2.44$. This yields
five more spectroscopic redshifts.

In total, we obtain 61 possible group members (see Fig.~\ref{f.group} yellow circles for the subset of objects closest to the light-weighted centre), and
identify a galaxy overdensity with peak significance level that is nearly 5 standard deviations above the background level at these redshifts
\citetext{Fig.~\ref{f.overdensity.a}, and see \citealp{helton_jades_2023}}. The resulting redshift distribution peaks at 2.27 (using the
input redshifts) or at 2.31 (using the median value from \beagle photometry-only fits); the
median uncertainty (zConf/2) is 0.06.
The location and extent of the overdensity are illustrated in Fig.~\ref{f.overdensity.a},
using a kernel density estimate (KDE) with a bandwidth (smoothing scale) of 1~cMpc.
The peak of the KDE density is located at
53.12262, $-27.80490$ which lies in projection only 3.5~arcsec away from our target and
5.4~arcsec away from the massive central.

Crucially, the photometry-only
selection does not include our main spectroscopic target; its lack of emission lines makes 
the photometric redshift quite uncertain.
Relaxing the redshift selection criteria until we include \targetidfull in the photometric
selection increases the noise in the overdensity determination. The fact that we would not have 
selected \targetidfull based on photometry alone underscores the importance of spectroscopy 
when studying the relation between environment and quenching: even with the depth of JADES,
the photometric-redshift selection at $z=2.3$ is biased towards galaxies with strong
photometric excess due to nebular-line emission, i.e., star-forming rather than quiescent
galaxies.
Searching for extended X-ray emission in the
catalogue of \citet{finoguenov_chandra_2015} does not yield any matches: their 4Ms data from
the Extended Chandra Deep Field South only reaches $z=1.6$.

We also explore a complementary selection method to identify members of an overdensity at
these redshifts. We search for possible quiescent objects detected in
F115W, with a break between F115W and F150W (flux ratio larger than 2.5, or 1~mag), and no strong
excess between F200W and F150W (ratio less than 20~per cent or 0.2~mag, to avoid high equivalent-width
\oiii emission mimicking the break). We apply this selection
only 10~arcsec from the central galaxy, where the density of members is expected to be
highest and, therefore, the fraction of contaminants lowest. This selection is highlighted 
by the white diamonds labelled a--d in Fig.~\ref{f.group}. By construction, we include our 
main target (a), but in addition we find three other galaxies which were not included in the
robust photometric selection. These are a relatively compact satellite 
(b), an extended low-surface-brightness galaxy close to the central (d) and an even
closer satellite clearly interacting with the central (c). The detection of
four objects is significant at the 3-\textsigma level: repeating the selection process within 
10~arcsec from 10,000 randomly chosen locations, we find three or less objects 99.9~per
cent of the time.

\section{Discussion}\label{s.discussion}

\subsection{Balmer or Lyman break?}\label{ss.z9}

For a redshift $z=9$ galaxy, the lack of emission lines could be explained by a
high escape fraction \citetext{e.g., \citealp{Zackrisson2017ApJ...836...78Z},
Bunker et~al., in~prep.}, but this seems at odds with the relatively flat
UV slope $\beta$ (e.g., Fig.~\ref{f.prosp.model}), because galaxies with high 
escape fractions tend to have steep UV slopes \citetext{e.g., 
\citealp{Zackrisson2017ApJ...836...78Z}, Topping et~al., in~prep.}. There is
now evidence for a few spectroscopically confirmed high-redshift galaxies 
with relatively high $\beta$
\citep[$-2.2<\beta<-1.8$;][]{Curtis_Lake2022arXiv221204568C}, but these galaxies 
are all at $z>10.5$, where the strongest nebular emission lines (\Hbeta and
\oiii) are all outside the wavelength range of NIRSpec. Unlike for $z>10.5$,
at $z=9$ both \Hbeta and \oiii are observable with NIRSpec,
therefore, if our galaxy were at $z=9$, the non detection of these lines would 
imply no ongoing star formation.
While galaxies with no rest-frame optical emission lines have indeed been found at 
redshifts $z=5\text{--}7$ \citep{Looser+2023, Strait2023}, these rapidly quenched 
galaxies also exhibit a Balmer break, for which there is no evidence in our target 
(at observed wavelength 3.6~\mum, Fig.~\ref{f.img.c}).

In contrast, at $z=2.34$, the lack of emission lines could be easily explained if 
the object was quiescent. SED modelling infers a stellar mass of $\mfinal$~\MSun; 
passive galaxies at these redshifts and in this mass range are indeed predicted 
(and abundant) according to theoretical models \citep{tacchella_efficiency_2018, donnari_activity_2019, donnari_quenched_2021, Dome2023}.
The extreme proximity of the target to a massive, spectroscopically confirmed galaxy at 
$z=2.349$ (\targetidcentral), the traces of interaction elongated in the direction of the
central galaxy (Fig.~\ref{f.img.a}), and the statistically significant presence of three other 
galaxies of similar colour within 10~arcsec all suggest this system is a satellite galaxy
at $z=2.34$.

The possibility of a mis-identified $z=9$ object remains (especially given
the high break measurement of $2.7\pm0.7$), but it would mean all 
other lines of evidence must be remarkable coincidences. In the following, we 
discuss the implications for the fiducial redshift interpretation.

\subsection{Massive galaxies in overdensities: nature or nurture?}

Current evidence for environment-driven quenching at high redshift focuses
on high-mass systems. A number of studies have confirmed that high-mass
galaxies in overdensities at $z=2\text{--}5$ tend to be more massive and
older than in the field \citep[e.g.,][]{
Lemaux_VIMOS_18, shimakawa_mahalo_2018, Shi_accelerated_21}. At redshifts
$z>3$ observational selection techniques could be biased due to the difficulty of distinguishing the
spectra of old stellar populations and dust-obscured star-forming systems
\citep[e.g.,][]{alberts_noble_2022}. However, at $z\approx2$, our methods are
sufficiently robust to confirm the presence of old, evolved systems
\citetext{via SED fitting, e.g., \citealp{cassata_constraining_2013} or with
the UVJ diagram, e.g., \citealp{ji_enviromental_2018}}. The overdensity we find
is no different than other overdensities at these redshifts: the three most 
massive galaxies all have high \Mstar and low  specific SFR -- two of them 
consistent with quiescence.
While the higher-than-average \Mstar is likely associated with the overdense environment,
the relatively old stellar populations and low sSFR may not be \textit{caused} by
environment -- at least not directly. Theoretically, we know that at masses above $10^{10}$~\MSun quenching may
occur due to internal mechanisms \citep[as suggested by simulations, which correctly 
predict a rising fraction of massive, quenched central galaxies at 
$z=2\text{--}3$;][]{donnari_quenched_2021}. Moreover, empirically, 
there exist a few massive, quiescent
galaxies at high redshift without obvious neighbours
\citep[e.g.,][]{carnall+2023c}. In principle, this could be due to
insufficient spectroscopy observations \citetext{see e.g., 
\citealp{williams_alma_2021, williams_alma_2022} for evidence of 
satellites around quiescent galaxies that were previously thought to be 
isolated} or to the presence of heavily dust-obscured companions 
\citep{schreiber_jekyll_2018}.
However, several lines of evidence suggest that the overabundance of 
red/quiescent massive galaxies in high-redshift protoclusters may be a 
consequence of higher average \Mstar compared to the field and of
mass-related quenching, which may be more efficient at higher
redshifts \citetext{\citealp{peng+2010, cassata_constraining_2013}; here `mass-related' quenching means any mechanism that scales with 
\Mstar, e.g. quenching from supermassive black-holes,
\citealp{bluck+2022}}.

\subsection{Environment-driven quenching at the low-mass end}

Unlike massive galaxies, lower-mass systems in the range $\Mstar = 
10^8\text{--}10^{10}$~\MSun are not expected to become quiescent from 
internal mechanisms. Indeed, numerical simulations predict a negligible fraction of
non-satellite galaxies in this mass range to be already
quiescent at $z=2$ \citep{donnari_quenched_2021}. Therefore, these systems 
represent our best opportunity for studying environment-driven quenching. For
example, \citet{ji_enviromental_2018} 
have used a sample of $\approx$600 colour-selected quiescent galaxies at redshifts 
$z=1.6\text{--}2.6$ and compared their spatial clustering to that of star-forming galaxies,
finding that quiescent galaxies are indeed more clustered, as expected from
environment-driven quenching.
However, when the depth of the observations 
is limited, the sample selection is inevitably biased toward star-forming galaxies.
As we have seen in \S~\ref{s.env}, a robust redshift selection necessarily removes the 
lowest-mass satellites -- precisely the objects where environment effects are 
predicted to have the highest impact.
\citet{ji_enviromental_2018} find that only about 4~per cent of their quiescent
galaxies are in the lowest-mass bin $10^9<\Mstar \lesssim 2\times10^9$~\MSun.
Using narrow-band photometry targeting \Halpha in the $z=2.2$ Spiderweb 
protocluster, \citet{shimakawa_mahalo_2018} find an increasing \Halpha equivalent
width with decreasing \Mstar, down to stellar masses of $10^9$~\MSun, i.e., in the
regime where up to 40~per cent of satellites are expected to be quiescent at $z=2$ 
\citep{donnari_quenched_2021}; this is contrary to what we expect for an unbiased
sample.
The galaxy we present here and its surrounding environment fit perfectly into the 
current theoretical framework.
We have three massive, evolved galaxies, surrounded by up to 58
satellites (based on robust photometric redshifts). Among the most nearby satellites
(within 10~arcsec) we find three additional satellite galaxies with Balmer (or 4000-\AA) 
breaks and low equivalent width emission lines -- possibly tracing galaxies with similar 
physical properties as our main target. Our target has relatively large
\re (for its \Mstar), S\'ersic index consistent with a disc ($n=1$;
\S~\ref{s.photometry}), and displays an inverse colour gradient with radius
(Fig.~\ref{f.colour}). These properties suggest a different evolutionary
path compared to the massive, quiescent galaxies in the same environment.
In particular, the inverse colour gradient is consistent with outside-in
quenching, which is interpreted as evidence for environment-induced
quiescence in local galaxies \citep{bluck+2020b}.

For \targetidfull, deep JADES spectroscopy enables us to accurately measure 
\Mstar and the quenching time. With $\Mstar = \mfinal$~\MSun, this galaxy is simply
too massive to have quenched as a result of the cosmic UV background
\citep{efstathiou_suppressing_1992, Ma2018MNRAS.478.1694M}.
The quenching time is \tquench~Gyr after the Big Bang, corresponding to a quenching redshift 
$z=$~\zquench. This measurement enables
us to trace environment-driven quenching well before the epoch of observation, showing that
environment started to play a decisive role very early in the history of the Universe, when
the Universe was only $\approx 2$~Gyr old.
Together with emerging evidence for early structures at $z=5\text{--}7$ 
\citep{Lemaux_VIMOS_18, laporte_a_dense_2022, brinch_cosmos2020_2023}, our findings
underscore the importance of environment with regards to interpreting the SFH of galaxies -- even at high 
redshifts.

A central element of our analysis is that spectroscopy enables us to accurately probe the
low-mass end of the satellite-galaxy distribution, which is essential to (1) identify
large-scale structures robustly and (2) derive environmental dependencies for the SFH of
galaxies.
Even though the depth of our observations is not easily achieved for the large samples 
required to fully characterise environment-driven quenching, environmental effects should 
still be measurable for masses $\Mstar=10^{9.5}\text{--}10^{10}$~\MSun, which are within 
reach of ground-based surveys like MOONRISE \citep{maiolino+2020} and of future, targeted 
surveys with \jwst.

\section{Summary and Conclusions}\label{s.summary}

We have presented deep \jwst/NIRCam and NIRSpec observations of \targetidfull, a compact quiescent galaxy at $z=2.34$,
identified through its spectral break at 1.25~\mum and the absence of emission lines.

More specifically, we summarise our findings in the following:

\begin{enumerate}
    \item We have used \forcepho to measure its light profile, finding
    $\re = 0.72\pm0.02$~kpc and $n=1.00\pm0.07$;
    our galaxy has a different structure than more massive galaxies at
    similar redshift, a possible sign of different evolutionary path.
    \item We have used full spectral modelling with \beagle and joint spectro-photometric modelling 
    with \prospector to measure its physical properties; we have found a stellar mass
    $\Mstar = \mfinal$~\MSun and negligible SFR in the last 100~Myr prior to observation,
    meaning the object is quiescent (Figs.~\ref{f.corners} and \ref{f.sfh}).
    \item We have estimated that the SFH peaked 500--1,000 Myr prior to observation (Fig.~\ref{f.sfh}),
    the mass-weighted age is 0.8--1.7~Gyr (dominated by systematic uncertainties),
    and the quenching redshift is \zquench, corresponding to a time when the
    Universe was only 2~Gyr old.
    \item PSF-matched photometry shows an inverse colour gradient with 
    radius (Fig.~\ref{f.colour}), consistent with the expectation from
    environment-driven quenching.
    \item We have identified a $\delta \sim 5$~\textsigma overdensity centred very near to the position of
    the target (Fig.~\ref{f.overdensity}). This consists of 61 targets with robustly
    determined photometric redshifts and spectroscopic redshifts from the literature.
    The three most massive galaxies in the overdensity ($\Mstar = 8\text{--}17\times10^{10}$~\MSun)
    all lie within 24~arcsec (0.66~cMpc) of the target (Fig.~\ref{f.group}),
    have photometric breaks consistent with the Balmer or 4000-\AA break and
    with evolved stellar populations at $z=2.34$. Due to the mass of these galaxies,
    their quiescence may be due to mass-related (internal) mechanisms, rather than
    environment.
    \item Of these three galaxies, the one closest to our target has a spectroscopic
    redshift $z=2.349$ and lies only 4~arcsec (35~pkpc) away in projection, at the
    centre of the overdensity.
    \item Based on the photometric selection alone, we would \textit{not} have selected 
    \targetidfull. Even with the depth of JADES, robust photometric selection of 
    overdensities at $z\approx2$ requires robust photometric redshift measurements to
    overcome background noise. This means overdensity selections are biased toward galaxies 
    with strong photometric excess due to nebular-line emission, i.e., star-forming rather than quiescent galaxies. This bias must be considered when studying environment-driven 
    quenching at high redshift.
    \item Within 10~arcsec from the central galaxy, we have identified three
    additional faint galaxies consistent with low SFR or quiescence, based on having
    photometry similar to our target.
    \item Summarising, \targetidfull is a low-mass, quiescent satellite galaxy close to its 
    central, making it the earliest spectroscopic evidence of environment-driven quenching
    to date.
\end{enumerate}

\begin{acknowledgements}

LS, FDE, RM, WB, TJL acknowledge support by the Science and Technology Facilities Council (STFC), by the ERC through Advanced Grant 695671 ``QUENCH'', and by the
UKRI Frontier Research grant RISEandFALL. RM also acknowledges funding from a research professorship from the Royal Society.
AJB and JC acknowledge funding from the ``FirstGalaxies'' Advanced Grant from the European Research Council (ERC) under the European Union's Horizon 2020 research and innovation programme (Grant agreement No. 789056).
SC acknowledges support by European Union's HE ERC Starting Grant No. 101040227 - WINGS.
ECL acknowledges support of an STFC Webb Fellowship (ST/W001438/1).
BDJ, BR, CNAW, DJE, JMH, KH, SA and ZJ acknowledge support from the JWST/NIRCam Science Team contract to the University of Arizona, NAS5-02015; DJE is also supported as a Simons Investigator.
BRP acknowledges support from the research project PID2021-127718NB-I00 of the Spanish Ministry of Science and Innovation/State Agency of Research (MICIN/AEI/ 10.13039/501100011033)
H{\"U} gratefully acknowledges support by the Isaac Newton Trust and by the Kavli Foundation through a Newton-Kavli Junior Fellowship.
The research of CCW is supported by NOIRLab, which is managed by the Association of Universities for Research in Astronomy (AURA) under a cooperative agreement with the National Science Foundation. The authors acknowledge the FRESCO team led by PI Pascal Oesch for developing their observing programme with a zero-exclusive-access period. 

This work was performed using resources provided by the Cambridge Service for Data Driven Discovery (CSD3) operated by the University of Cambridge Research Computing Service (\href{www.csd3.cam.ac.uk}{www.csd3.cam.ac.uk}), provided by Dell EMC and Intel using
Tier-2 funding from the Engineering and Physical Sciences Research Council (capital grant EP/T022159/1), and DiRAC funding from the Science and Technology Facilities Council (\href{www.dirac.ac.uk}{www.dirac.ac.uk}).

This work made extensive use of the freely available
\href{http://www.debian.org}{Debian GNU/Linux} operative system. We used the
\href{http://www.python.org}{Python} programming language
\citep{vanrossum1995}, maintained and distributed by the Python Software
Foundation. We further acknowledge direct use of
{\sc \href{https://pypi.org/project/astropy/}{astropy}} \citep{astropyco+2013},
\beagle \citep{chevallard_beagle_2016},
{\sc \href{https://pypi.org/project/matplotlib/}{matplotlib}} \citep{hunter2007},
{\sc \href{https://pypi.org/project/numpy/}{numpy}} \citep{harris+2020},
\href{https://github.com/bd-j/prospector}{\prospector} \citep{johnson+2021},
{\sc \href{https://scikit-learn.org/stable/index.html}{scikit-learn}} \citep{scikit-learn},
{\sc \href{https://pypi.org/project/scipy/}{scipy}} \citep{jones+2001},
{\sc \href{https://pypi.org/project/smplotlib/}{smplotlib}} \citep{smplotlib}
and {\sc \href{http://www.star.bris.ac.uk/~mbt/topcat/}{topcat}} \citep{taylor2005}.

\end{acknowledgements}

%
%

\bibliographystyle{config/aa}
\bibliography{Bib_TL}

\begin{appendix}

\section{Photometric fitting}\label{a.photometry}

\begin{figure}
   \centering
   \includegraphics[width=1.1\textwidth,angle=-90]{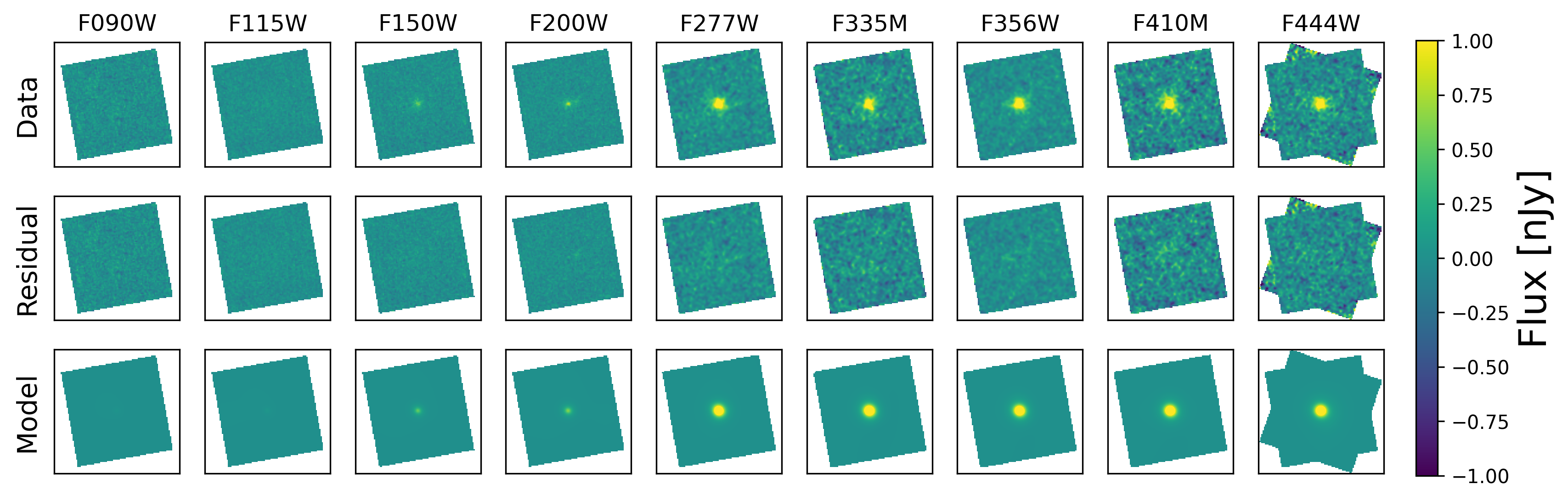}
   \caption{Model \forcepho photometry, illustrating the data (top row),
   the S\'ersic model (bottom row) and their difference (middle row). The
   residuals show no clear evidence for a second component, except for F200W,
   where there is an extended feature which we already noted in Fig.~\ref{f.img.a};
   note how this feature is not present in the other bands, as discussed in
   \S~\ref{s.photometry}. The images are cut along the native pixel grid of
   the NIRCam detectors; \forcepho fits the individual exposures to avoid
   the correlated uncertainties due to resampling.
   }\label{f.forcepho}
\end{figure}

To model the light distribution of our target, we use \forcepho (Johnson et~al.,
in~prep.), following the approach of \citet{baker_insideout_2023}. We assume a
single-component \citet{sersic1968} light profile where the flux in each band is allowed to 
vary, but the structural parameters are independent of wavelength (half-light
semi-major axis \re, position angle and axis ratio $q$).
We choose here to forward model the light distribution in the NIRCam images in
the filters F090W, F115W, F150W, F200W, F277W, F335M, F356W, F410M and F444W
(Fig.~\ref{f.forcepho}). \forcepho fits the S\'ersic profile to all
individual exposures simultaneously, accounting for the PSF in each band. The
model is optimised using a Markov Chain Monte Carlo algorithm, which enables us to 
estimate the uncertainties on the model parameters as well as their covariances.
Running on the individual exposures also avoids introducing correlated measurement
uncertainties from resampling the images onto a common grid.

The marginalised posterior distribution gives $\re=0.087\pm0.002$~arcsec,
$n=1.00\pm0.07$, and $q=0.91\pm0.04$. The half-light radius is lighly larger than 
the full-width at half maximum of the empirical NIRCam PSF at 1.5 and 2.0~\mum,
but clearly smaller than the PSF at longer wavelenghts \citep[e.g.,][]{ji_jadesjems_2023}.
With the adopted cosmology, the effective
radius corresponds to $\re=0.72\pm0.02$~kpc. There is currently no census of
the size of quiescent galaxies at redshift $z=2.34$ in the same mass range as
our target \citep[e.g.,][]{vanderwel+2014}. At these redshifts, \re=0.7~kpc is the 
mean \re of quiescent galaxies 30 times more massive than our target
\citep[$\Mstar=3\times10^{10}$~\MSun,][]{vanderwel+2014}. Comparing this galaxy
to the central, the two galaxies have the same \re (within the uncertainties),
but the central has much more concentrated light profile
\citep[$n=5.7\pm0.1$;][]{cassata_constraining_2013} and the mass ratio is over
one hundred.
If \targetidfull was representative of quiescent, low-mass satellite galaxies at
$z\approx2$, these galaxies may break away from the mass-size relation of more 
massive quiescent systems -- which could be a litmus test of environment
quenching.

\section{SED of the central galaxy}\label{a.central}

The central galaxy \targetidcentral is bright ($H$-band Kron magnitude
22.1~mag) and has a spectroscopic redshift of $z=2.349\pm0.001$
\citep[determined from \lya emission;][their object C-3119]{wuyts_optical_2009}. The galaxy was also studied by \citet{cassata_constraining_2013}, who report a photometric
redshift $z_\mathrm{phot}=2.326$ based on photometric SED fitting.

\begin{figure}
   \centering
   \includegraphics[width=\columnwidth]{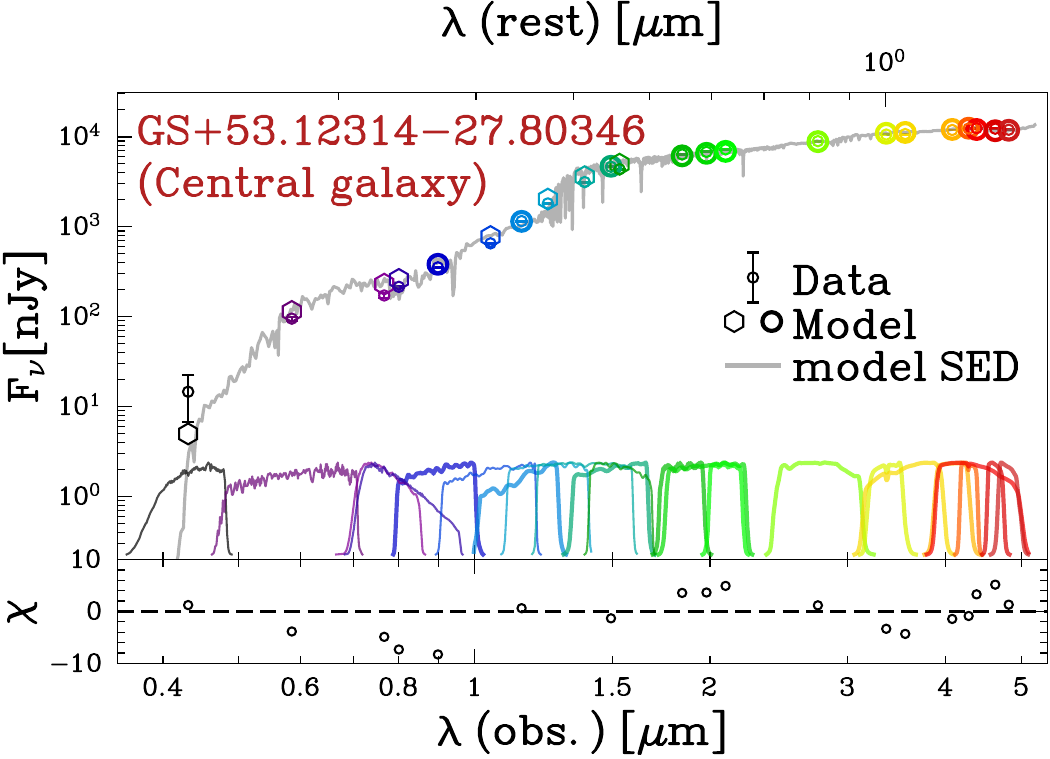}
   \caption{Observed SED of the central galaxy (empty circles with errorbars) and maximum a-posteriori model from \prospector (larger hexagons/circles mark \hst/\jwst photometry).
   The data shows no evidence for photometric excess in any of the filters, which is captured
   by the \prospector model (grey line).
   We imposed a floor on the flux uncertainties to be larger than 1~per cent.
   }\label{f.central}
\end{figure}

For our re-analysis, we used \prospector with the same setup as for the 
main target \citetext{but without spectroscopy, because the spectrum in
\citealp{wuyts_optical_2009} is not flux calibrated}. We use Kron
photometry (to capture the total flux) and we add an artificial floor 
to the flux uncertainties of 1~per cent (our measurements have a
nominal uncertainty as low as 0.3~per cent).
The observed SED and the \prospector model are shown in 
Fig.~\ref{f.central}. Given the lack of photometric excess in the 
medium- and broad-band filters, there seems to be little 
or no room for strong nebular emission in the rest-frame optical. The 
maximum a-posteriori model (grey line) has no \lya emission, in
disagreement with the detection reported by \citet{wuyts_optical_2009}.
However, the \hst/ACS flux in F435W seems under-estimated, a
possible indication of \lya emission. Trying to `force' an emission-line
solution by artificially up-weighting the F435W photometry does not 
change our results: the number and quality of the rest-frame optical 
measurements leaves no room for any nebular emission.
Repeating the analysis with \beagle instead of \prospector, these results
are unchanged.
The galaxy is not detected in X rays.

We conclude that our model is uncertain due to possible AGN/interloper
contamination, while both \prospector and \beagle photometry-only
models do not show any  evidence for line emission. 
Despite these uncertainties, the final stellar mass is within a factor
of two from the value reported in \citet{cassata_constraining_2013}.

\end{appendix}

\end{document}

%% file: authors.tex
\author{
\orcid{Lester Sandles}{0000-0001-9276-7062}
\inst{\hyperlink{aff1}{1}}\fnmsep\inst{\hyperlink{aff2}{2}}
\and
\orcid{Francesco D'Eugenio}{0000-0003-2388-8172}
\inst{\hyperlink{aff1}{1}}\fnmsep\inst{\hyperlink{aff2}{2}}
\and
\orcid{Jakob M. Helton}{0000-0003-4337-6211}
\inst{\hyperlink{aff3}{3}}
\and
\orcid{Roberto Maiolino}{0000-0002-4985-3819}
\inst{\hyperlink{aff1}{1}}\fnmsep\inst{\hyperlink{aff2}{2}}\fnmsep\inst{\hyperlink{aff4}{4}}
\and
\orcid{Kevin Hainline}{0000-0003-4565-8239}
\inst{\hyperlink{aff3}{3}}
\and
\orcid{William M. Baker}{0000-0003-0215-1104}
\inst{\hyperlink{aff1}{1}}\fnmsep\inst{\hyperlink{aff2}{2}}
\and
\orcid{Christina C. Williams}{0000-0003-2919-7495}
\inst{\hyperlink{aff5}{5}}
\and
\orcid{Stacey Alberts}{0000-0002-8909-8782}
\inst{\hyperlink{aff3}{3}}
\and
\orcid{Andrew J.\ Bunker }{0000-0002-8651-9879}
\inst{\hyperlink{aff6}{6}}
\and
\orcid{Stefano Carniani}{0000-0002-6719-380X}
\inst{\hyperlink{aff7}{7}}
\and
\orcid{Stephane Charlot}{0000-0003-3458-2275}
\inst{\hyperlink{aff8}{8}}
\and
\orcid{Jacopo Chevallard}{0000-0002-7636-0534}
\inst{\hyperlink{aff6}{6}}
\and
\orcid{Mirko Curti}{0000-0002-2678-2560}
\inst{\hyperlink{aff9}{9}}\fnmsep\inst{\hyperlink{aff1}{1}}\fnmsep\inst{\hyperlink{aff2}{2}}
\and
\orcid{Emma Curtis-Lake}{0000-0002-9551-0534}
\inst{\hyperlink{aff10}{10}}
\and
\orcid{Daniel J.\ Eisenstein}{0000-0002-2929-3121}
\inst{\hyperlink{aff11}{11}}
\and
\orcid{Zhiyuan Ji}{0000-0001-7673-2257}
\inst{\hyperlink{aff3}{3}}
\and
\orcid{Benjamin D.\ Johnson}{0000-0002-9280-7594}
\inst{\hyperlink{aff11}{11}}
\and
\orcid{Tobias J. Looser}{0000-0002-3642-2446}
\inst{\hyperlink{aff1}{1}}\fnmsep\inst{\hyperlink{aff2}{2}}
\and
\orcid{Tim Rawle}{0000-0002-7028-5588}
\inst{\hyperlink{aff12}{12}}
\and
\orcid{Brant Robertson}{0000-0002-4271-0364}
\inst{\hyperlink{aff13}{13}}
\and
\orcid{Bruno Rodríguez Del Pino}{0000-0001-5171-3930}
\inst{\hyperlink{aff14}{14}}
\and
\orcid{Sandro Tacchella}{0000-0002-8224-4505}
\inst{\hyperlink{aff1}{1}}\fnmsep\inst{\hyperlink{aff2}{2}}
\and
\orcid{Hannah \"Ubler}{0000-0003-4891-0794}
\inst{\hyperlink{aff1}{1}}\fnmsep\inst{\hyperlink{aff2}{2}}
\and
\orcid{Christopher N. A. Willmer}{0000-0001-9262-9997}
\inst{\hyperlink{aff3}{3}}
\and
\orcid{Chris Willott}{0000-0002-4201-7367}
\inst{\hyperlink{aff15}{15}}
}

\institute{
\hypertarget{aff1}{Kavli Institute for Cosmology, University of Cambridge, Madingley Road, Cambridge, CB3 0HA, UK}\\
\and
\hypertarget{aff2}{Cavendish Laboratory, University of Cambridge, 19 JJ Thomson Avenue, Cambridge, CB3 0HE, UK}\\
\and
\hypertarget{aff3}{Steward Observatory, University of Arizona, 933 North Cherry Avenue, Tucson, AZ 85721, USA}\\
\and
\hypertarget{aff4}{Department of Physics and Astronomy, University College London, Gower Street, London WC1E 6BT, UK}\\
\and
\hypertarget{aff5}{NSF’s National Optical-Infrared Astronomy Research Laboratory, 950 North Cherry Avenue, Tucson, AZ 85719, USA}\\
\and
\hypertarget{aff6}{Department of Physics, University of Oxford, Denys Wilkinson Building, Keble Road, Oxford OX1 3RH, UK}\\
\and
\hypertarget{aff7}{Scuola Normale Superiore, Piazza dei Cavalieri 7, I-56126 Pisa, Italy}\\
\and
\hypertarget{aff8}{Sorbonne Universit\'e, CNRS, UMR 7095, Institut d'Astrophysique de Paris, 98 bis bd Arago, 75014 Paris, France}\\
\and
\hypertarget{aff9}{European Southern Observatory, Karl-Schwarzschild-Strasse 2, 85748 Garching, Germany}\\
\and
\hypertarget{aff10}{Centre for Astrophysics Research, Department of Physics, Astronomy and Mathematics, University of Hertfordshire, Hatfield AL10 9AB, UK}\\
\and
\hypertarget{aff11}{Center for Astrophysics $|$ Harvard \& Smithsonian, 60 Garden St., Cambridge MA 02138 USA}\\
\and
\hypertarget{aff12}{European Space Agency (ESA), European Space Astronomy Centre (ESAC), Camino Bajo del Castillo s/n, 28692 Villafranca del Castillo, Madrid, Spain}\\
\and
\hypertarget{aff13}{Department of Astronomy and Astrophysics, University of California, Santa Cruz, 1156 High Street, Santa Cruz, CA 95064, USA}\\
\and
\hypertarget{aff14}{Centro de Astrobiolog\'ia (CAB), CSIC–INTA, Cra. de Ajalvir Km.~4, 28850- Torrej\'on de Ardoz, Madrid, Spain}\\
\and
\hypertarget{aff15}{NRC Herzberg, 5071 West Saanich Rd, Victoria, BC V9E 2E7, Canada}\\
}

%% file: model_table.tex
\begin{table*}
    \begin{center}
    \caption{Parameters and associated priors set in \beagle (left) and \prospector (right).}
    \label{tab.priors}
    \begin{tabular}{c c c c }
        \toprule
        \multicolumn{2}{c}{\beagle} & \multicolumn{2}{c}{\prospector} \\
        Parameter \& Symbol  & Prior & Parameter \& Symbol & Prior \\
        \midrule

        Redshift                         $z$                  & $\mathcal{N}(2.34, 0.5^2)$                                  & " & $\mathrm{Unif.} \in [1.84, 2.84]$\\ 
        Mass formed                      $\log(\Mstar/\MSun)$ & $\mathrm{Unif.} \in [6, 12]$                              & " & $\mathrm{Unif.} \in [7, 11]$ \\ %
        Stellar metallicity              $\log(Z_\star/\ZSun)$      & $\mathrm{Unif.} \in [-2.2, 0.4]$                          & " & $\mathrm{Unif.} \in [-2, 0.19]$ \\ 
        \\
                                         
        Age of oldest stars              $\log(t / \mathrm{yr})$ & $\mathrm{Unif.} \in [6, 13^{\dagger}]$                           & Fixed time bins & 8 $\log$-spaced bins \\ 
        Timescale of SFH                 $\log(\tau / \mathrm{yr})$ & $\mathrm{Unif.} \in [6, 12]$                        & Ratio of $\log$ SFR between bins & Student's $t(0, 0.3, 2)$ \\ 
                                         
        SFR of last bin               $\log(\mathrm{SFR}_\mathrm{const} / \MSun\peryr)$ & $\mathrm{Unif.} \in [-4, 4]$ & -- & -- \\ 
        Duration of const.               $\log(t_{\mathrm{const}} / \mathrm{yr})$ & $\mathrm{Unif.} \in [7, 9]$           & -- & -- \\ 
                                         
        \\
        
        Total $V$-band att.              $\tauV$ & $\exp(-\tauV)$, $\tauV \in [0, 6]$ & $V$-band att. in diff. ISM \tvdiff & $\mathcal{N}(0.3, 1) \in [0, 2]$ \\ 
        Fraction of $\tauV$ in diff. ISM $\mud$ & Fixed 0.4 & Ratio of diff. to birth-cloud att. $f_\mathrm{d}^{\ddagger}$ & $\mathcal{N}(1, 0.3) \in [0, 2]$ \\ 
        --                                      &    --     & Dust power-law offset index        $n$  & $\mathrm{Unif.} \in [-1, 0.2]$ \\ 
        
        \\
        
        Ionisation parameter             $\logUs$ & $\mathrm{Unif.} \in [-4, -1]$ & " & " \\ 
        Dust-to-metal mass ratio         $\xid$ & $\mathrm{Unif.} \in [0.1, 0.5]$ & -- & -- \\ 
        ISM metallicity                  $\log(Z_{\mathrm{ISM}}/\ZSun)$ & $\mathrm{Unif.} \in [-2.2, 0.4]$ & " & $\mathrm{Unif.} \in [-2, 0.5]$ \\ 
        
        \bottomrule
    \end{tabular}
    \end{center}
    For \prospector, `"' indicates that the meaning and/or prior of that parameter is the same as for \beagle. `--' indicates a parameter
    that is present in one model but not the other.
    $\mathcal{N}(a, b)$ is the Normal distribution with mean $a$ and standard deviation $b$; when an interval is also
    specified next to the Normal, (e.g., $\mathcal{N}(a, b) \in [c, d]$), this indicates a Normal distribution clipped
    between $c$ and $d$. The Student's distribution $t(a, b, c)$ has mean $a$, standard
    deviation $b$ and $c$ degrees of freedom.

    $\dagger$ In practice, \beagle will not allow the age of the oldest stars to be greater than the time between $z=20$ and the sampled redshift.
    
    $\ddagger$ To first order, one can relate the \beagle and \prospector dust-attenuation models with $\mud = (1+f_\mathrm{d})^{-1}$.
\end{table*}

%% file: posterior_table.tex
\begin{table}
    \begin{center}
    \caption{Posterior medians from \beagle and \prospector fits (with 16th and 84th percentiles) for the parameters shown in Fig.~\ref{f.corners}.}
    \label{tab.posteriors}
    \begin{tabular}{c c c }
        \toprule
        Parameter  & \beagle & \prospector  \\
        \midrule


$\log(\,M_\star \,/\, \mathrm{M}_\odot)$ & $8.98^{+0.07\,\,\,\dagger}_{-0.06}$ & $8.91^{+0.05}_{-0.06}$ \\ \\
$\log(\,Z \,/\, \mathrm{Z}_\odot)$ & $-1.2^{+0.3}_{-0.3}$ & $-1.7^{+0.3}_{-0.2}$ \\ \\
$\log(\,\mathrm{Age} \,/\, \mathrm{Gyr}\,)$ & $0.8^{+0.3}_{-0.2}$ & $1.7^{+0.3}_{-0.3}$ \\ \\
$\log(\,\mathrm{SFR}_\mathrm{100} \,/\, \mathrm{M}_\odot \, \mathrm{yr}^{-1}\,)$ & $-2.8^{+0.8\,\,\,\dagger}_{-0.7}$ & $-2.8^{+1.0}_{-1.5}$ \\ \\
$A_{V}$ & $0.6^{+0.2}_{-0.2}$ & $0.3^{+0.1}_{-0.1}$ \\

        \bottomrule
    \end{tabular}
    \end{center}

    $\dagger$ Including an additional $\log(1.54)$ and $\log(1.05)$ to account for the \prospector-derived aperture correction and the cosmology parameters used during the \beagle fit, respectively. 
    
\end{table}